%% file: Main_Document_updated.tex
\documentclass[twoside,BCOR=12mm]{sty/idcthesis}
\IDCThesisOptions{language=en,onehalfspacing=true,linkcolor=black!50!blue,fixfloatplacement=true}

\usepackage[latin1]{inputenc}
\usepackage{microtype}
\graphicspath{{graphics/}}
\usepackage{subfig}
\usepackage{amsmath,graphicx,cite}
\usepackage{algorithmic}
\usepackage{framed}

\usepackage[numbers,sort&compress]{natbib}

\usepackage{notoccite}

\usepackage{glossaries}
\usepackage{acronym}
\makeglossaries

\input{Macros}

\setupthesis{Master Thesis}
  {Multiuser Scheduling for Simultaneous Wireless Information and Power Transfer Systems}
  {Maryna Chynonova}
  {LaTeX-Vorlage, Hinweise zu LaTeX, Nomenklatur}
  {\today}

\begin{document}
\begin{spacing}{1}
    \KOMAoptions{cleardoublepage=empty}
    \pagestyle{empty} \pagenumbering{roman} \setcounter{page}{1}
    \input{0_Title}

    \inputdecl
    \tableofcontents
\end{spacing}
\let\cleardoublepage\clearpage
\addchap{\abstractname}
\input{0_Abstract}
\begin{spacing}{1}
   \addchap{\glossaryname}
   \input{0_Glossary}
     \listoffigures
\end{spacing}
\newcommand{\Un}{\text{U}}
\newcommand{\Sa}{\mathcal{S}_{\text{a}}}
\newcommand{\argorder}{\operatornamewithlimits{\text{arg order}}}
\newcommand{\argorderred}{\operatornamewithlimits{{\color{red}argorder}}}
\newcommand{\argminimum}{\operatornamewithlimits{\text{argmin}}}
\newcommand{\argmaximum}{\operatornamewithlimits{{\color{red}argmax}}}

\newtheorem{Thm}{Theorem}
\newtheorem{Prob}{Problem}
\newtheorem*{remark}{Remark}

\newcommand{\Na}{N_{A}}
\newcommand{\Nai}{$N_{A}$}
\newcommand{\Pobs}{P_{obs}(t)}
\newcommand{\Pobsi}{$P_{obs}(t)$}
\newcommand{\WRD}{\hat{W}_{R}}
\newcommand{\WRT}{\tilde{W}_{R}}
\newcommand{\OD}{\hat{W}}
\newcommand{\WRID}{\hat{W}_{R_{i}}}
\newcommand{\WRIT}{\tilde{W}_{R_{i}}}
\newcommand{\wrid}[1]{\hat{W}_{R_{#1}}}
\newcommand{\writ}[1]{\tilde{W}_{R_{#1}}}
\newcommand{\ExNumObs}[2]{\overline{N_{Aobs}}[#2] \big| {W[#2]=#1}}
\newcommand{\mOne}[1]{m_{1}[#1]}
\newcommand{\mZero}[1]{m_{0}[#1]}
\newcommand{\erf}{erf}
\renewcommand*{\figureformat}{%
  \figurename~\thefigure%
}

\chapter{Introduction}
\label{Introduction}
\pagenumbering{arabic} \setcounter{page}{1}

\section{Energy Harvesting in Wireless Networks}
Over the past decades, battery-powered devices have been deployed in many wireless communication networks. However, since batteries have limited energy storage capacity and their replacement can be costly or even infeasible, harvesting energy from the environment provides a viable solution for prolonging the network lifetime. Although conventional natural energy resources, such as solar and wind energy, are perpetual, they are   weather-dependent and location-dependent, which may not suitable for mobile communication devices. Alternatively, background radio frequency (RF) signals from ambient transmitters are also an abundant source of energy for energy harvesting (EH). Unlike the natural energy sources, RF energy is weather-independent  and can be available on demand. Nowadays, EH circuits are able to harvest microwatt to milliwatt of power over the range of several meters for a transmit power of $1$ Watt and a carrier frequency less than $1$ GHz  \cite{Powercast}. Thus, RF energy can be a viable energy source for devices with low-power consumption, e.g. wireless sensors \cite{Krikidis2014,Ding2014}. Moreover, RF EH provides the possibility for simultaneous wireless information and power transfer (SWIPT) since RF signals carry both information and energy \cite{Varshney2010,Grover2008}.

Currently, there are two main research directions on RF EH communications. The first direction studies the resource allocation algorithm design for SWIPT.  A fundamental tradeoff between information transfer rate and energy transfer rate was investigated in \cite{Zhang2013} under the assumption that the receiver is able to decode information and
extract power from the same received signal.  However, current practical circuits that harvest energy from RF signals are not yet able to decode the carried information directly from the same signal in general, i.e., the signal that is used for EH, cannot be reused for information decoding\footnote{We note that it is possible to decode the carried information directly from the same signal if non-coherent modulation schemes are adopted, e.g., on-off keying, in which information is carried on the energy level of the carrier signal.} \cite{Zhou2013}. Consequently, a power splitting receiver was proposed in \cite{Zhang2013} and \cite{Zhou2013} for facilitating simultaneous information decoding (ID) and EH. This receiver can be used for RF signals resulting from any modulation technique.

\begin{figure}[!h]
\centering
\includegraphics[width=4 in]{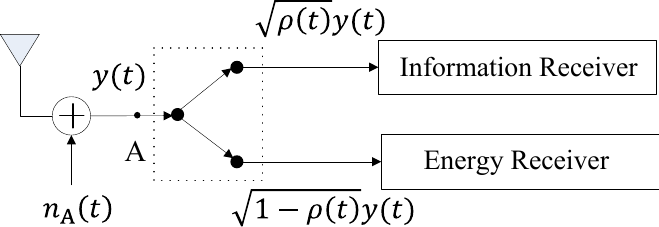}
\caption{Power splitting receiver.}
\label{fig:Fig3}
\end{figure}

A power splitting receiver  \cite{Zhou2013}--\nocite{CN:Kwan_globecom2013,CN:Eurosip_SWIPT,CN:ICC_WIPT_Kwan,JR:Kwan_secure_imperfect}\cite{JR:WIPT_fullpaper} consists of a conventional energy receiver and a conventional information receiver.
In particular, it splits the received signal into two power streams with power splitting ratios $1-\rho(t)$ and $\rho(t)$ at time instant $t$, cf. Figure \ref{fig:Fig3}, for harvesting energy and decoding the modulated  information, respectively.  Specifically, the power splitting unit is installed in the analog front-end of the receiver and is assumed to be a perfect passive analog device; it does not introduce any extra power gain, i.e., $0\le \rho(t)\le 1$, or noise to the received signal. In the extreme case, only ID is performed when $\rho(t)=1$ and only EH is done when $\rho(t)=0$, i.e., the receiver switches in time between the two modes. A receiver with $\rho (t) \in \{ 0, 1\} $ is referred to as a time-switching receiver. Besides, the power splitting receivers also generalize the case of  separated receivers in the literature \cite{CN:WCNC_WIPT}--\nocite{CN:kwan_vicky,CN:Kwan_PIMRC2013,CN:Multicast_SWIPT,CN:Kwan_globecom2014,CN:Maryna_2015,CN:tao_2015,
JR:MOOP_SWIPT,CN:Kwan_globecom2013,JR:rui_zhang,JR:Kwan_SEC_DAS,CN:PHY_SEC_max_min}\cite{Xu2013}.

Multiuser SWIPT systems have recently drawn significant attention in academia.  In \cite{JR:WIPT_fullpaper}, an orthogonal frequency division multiple access (OFDMA) system with SWIPT was considered. The authors designed a scheduling algorithm for the maximization of the energy efficiency of data transmission (bits/Joule delivered to the receivers) for a minimum required sum rate and a minimum required energy harvested by the users.  In \cite{JR:Kwan_secure_imperfect} and \cite{CN:kwan_vicky},  beamforming design was studied for power efficient  and secure SWIPT networks with imperfect channel state information (CSI) and perfect CSI,  respectively. In \cite{Xu2013},  beamformers were optimized for the maximization of the sum harvested energy under the minimum required signal-to-interference-plus-noise ratio constraints for multiple information receivers. Multiuser multiple input multiple output (MIMO) SWIPT systems were studied for the broadcast channel in \cite{Zhang2013} and for the interference channel in \cite{Park2013}.  Nevertheless, multiuser scheduling, which exploits multiuser diversity for improving the system performance of multiuser systems, has not
been considered in  \cite{Krikidis2014}--\cite{CN:PHY_SEC_max_min}. Recently,  simple suboptimal order-based schemes were proposed to balance the tradeoff between the users' ergodic achievable rates and their average amounts of harvested energy in \cite{Morsi2014}.  However, the  scheduling schemes proposed in \cite{Morsi2014} are unable to guarantee quality of service (QoS) with respect to the minimum energy transfer.  In fact, optimal multiuser scheduling schemes that guarantee a long-term minimum harvested energy for SWIPT systems have not been considered in the literature  so far.

\begin{figure}[!h]
\centering
\includegraphics[width=3.5 in]{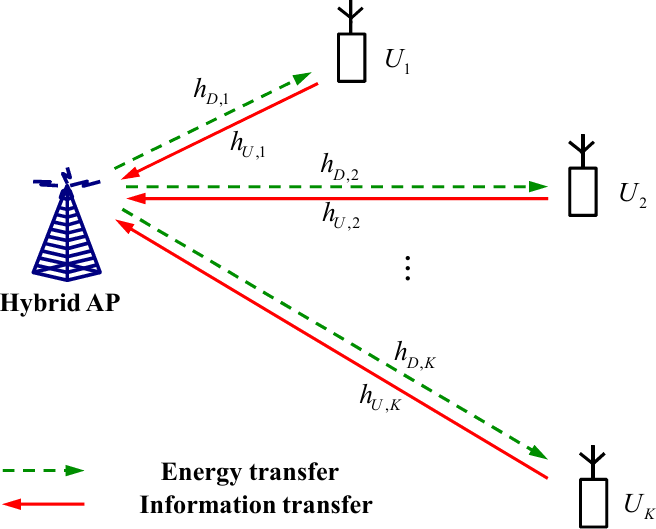}
\caption{A general wireless powered communication network (WPCN) \cite{LiuZahng2014}.}
\label{fig:WPCN}
\end{figure}

Another research direction focuses on wireless powered communication networks (WPCNs), where the wireless terminals in the network communicate using the energy harvested from wireless power transfer (WPT) as shown in Figure \ref{fig:WPCN} \cite{LiuZahng2014}.

In \cite{Huang2014}, the authors studied a wireless powered cellular network in which dedicated power-beacons were used in the cellular network to transfer wireless energy to mobile terminals. In \cite{Shi2011}, a wireless powered sensor network was investigated, where a mobile charging vehicle moving in the network was employed as the energy transmitter to wirelessly power sensor nodes. In \cite{Lee2013}, a network architecture enabling secondary users to harvest energy as well as reuse the spectrum of primary users in the cognitive radio network was proposed. In \cite{Ju2014}, the authors studied a typical WPCN model and proposed a harvest-then-transmit protocol, in which a hybrid access point (AP) coordinated WPT in the downlink (DL) and wireless information transfer (WIT) in the uplink (UL) for a set of distributed users. The authors proposed a protocol for sum-throughput maximization and enhanced it by considering a fair rate allocation among different users. In \cite{LiuZahng2014}, this system was extended to the case when the AP has multiple antennas. To improve the performance in wireless communication systems with EH, user cooperation was suggested and studied in \cite{Huang2013, Gurakan2012, JuZahng2014}. In \cite{ZahngJu2014}, full duplex technique was applied to a WPCN for further throughput improvement. Also, a new transmission protocol, enabling efficient simultaneous DL  WPT  and UL WIT over the same bandwidth, was proposed.

\section{Overview of the Thesis}
In this thesis, we investigate a SWIPT system with one AP and multiple users. We consider DL transmission with one user receives information in each time slot, while the remaining users opportunistically harvest the ambient RF energy. We design optimal scheduling algorithms that maximize the long-term average  system throughput under different fairness requirements, such as  proportional fairness and equal throughput fairness.  We show that with fairness considerations, the feasible trade-off regions of achievable data rate and harvested energy (R-E) for the proposed schemes decrease due to the lost of degrees of freedom in resource allocation.  Moreover, we study joint user selection and power allocation for the considered SWIPT systems. We propose an optimal resource allocation algorithm and reveal the R-E region of this scheme. In particular,  we show that joint power allocation and user scheduling is a more efficient way to enlarge the  feasible trade-off region  which improves the system performance in terms of achievable data rate and harvested energy.

The remainder of the thesis is organized as follows. In Section 2.1, we introduce the SWIPT system model. Sections 2.2 - 2.4 investigate the maximum throughput scheduling, proportional fair scheduling, and equal throughput scheduling schemes, respectively. For each scheduling scheme, we design the optimal user selection policy and verify its performance via simulation by comparing it with a baseline scheduling scheme. In Chapter 3, we consider joint user selection and power allocation for SWIPT systems and provide an optimal resource allocation algorithm. Finally, Chapter 4 concludes this thesis.

\chapter{Optimum Online Multiuser Scheduling}
In this chapter, we introduce our system model which consists of one AP and multiple users.  We design optimal scheduling algorithms that maximize the long-term average  system throughput under different fairness requirements, such as  proportional fairness and equal throughput fairness. In particular, the algorithm designs are formulated as  non-convex optimization problems which take into account the minimum required  average sum harvested energy in the system.

\section{System Model}
\begin{figure}[!h]
\centering
\includegraphics[scale=0.4]{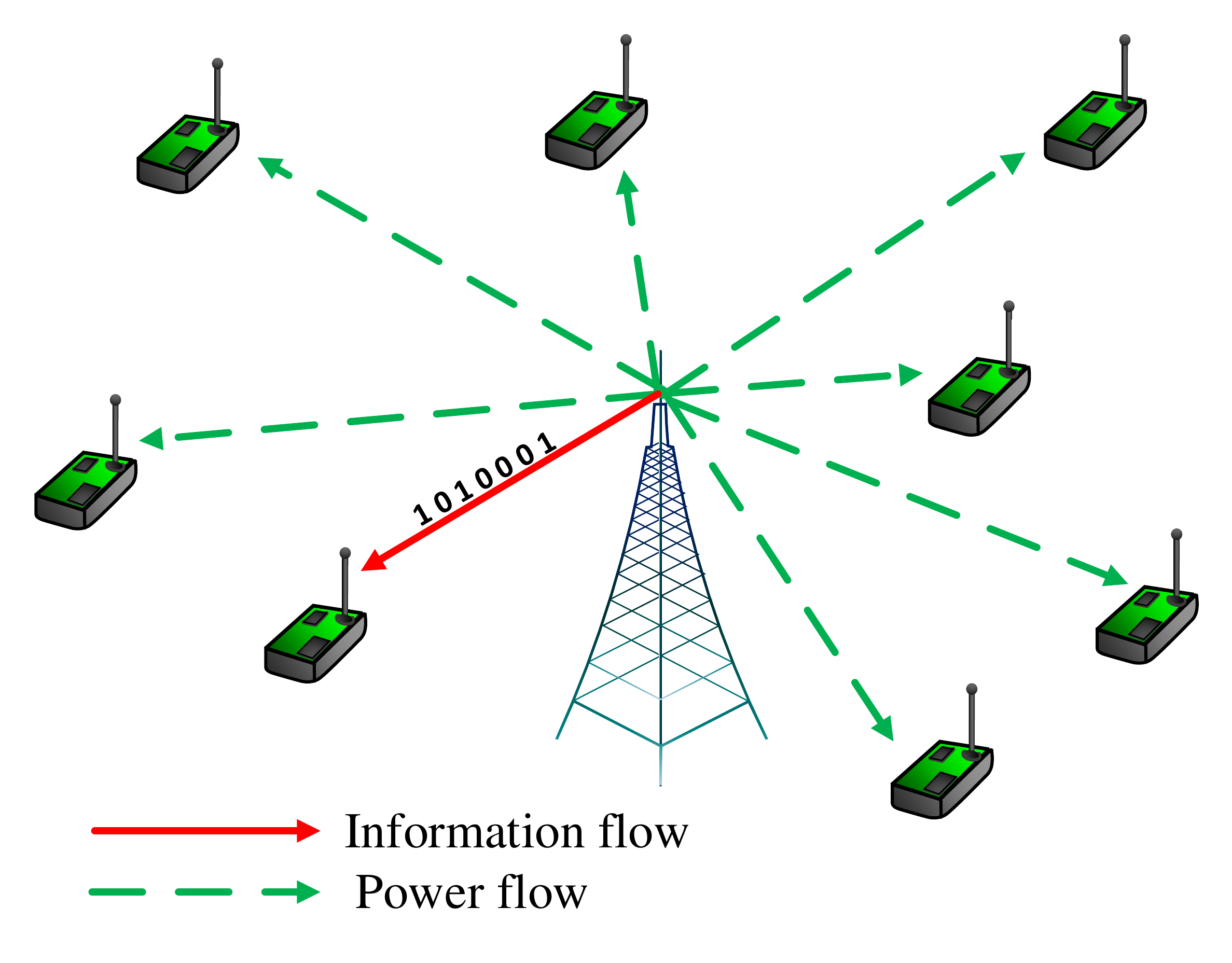}
\caption{SWIPT system model.}
\label{fig:Fig4}
\end{figure}

We consider a SWIPT system that consists of one access point (AP) with a fixed power supply and $N$ battery-powered user terminals (UTs), see Figure \ref{fig:Fig4}. The AP and the UTs are equipped with single antennas. Besides, we adopt time-switching receivers at the UTs \cite{Krikidis2014}  to ensure low hardware
complexity.

We study the user scheduling for DL transmission. We assume that the transmission is divided into $T$ time slots and  in each time slot perfect CSI is available at the AP. Also, the data buffer for the users at the AP is always full such that enough data packets are available for transmission for every scheduled UT. In each time slot, the AP schedules one user for ID, while the remaining users opportunistically harvest energy\footnote{We consider a unit-length time slot, hence the terms ``power" and ``energy" can be used interchangeably.} from the received signal.  We assume block fading channels. In particular, the channels remain constant during a time slot and change independently over different time slots. Besides, the users are physically separated from one another such that they experience independent fading.

 Furthermore, we adopt the  EH receiver model  from \cite{Zhou2013}. The RF energy harvested by user $n \in \{ 1, \ldots, N\}$ in time slot $i\in\{1,\ldots,T\}$ is given by
\begin{equation}
Q_n(i) = \xi_n P h_n(i),
\end{equation}
where $P$ is the constant AP transmit power, $0 \leq \xi_n \leq 1$ is the RF-to-direct-current (DC) conversion efficiency\footnote{(For currently available RF energy harvesters, the energy conversion efficiency can reach up to $0.7$ \cite{Powercast})} of the EH receiver of user $n$,  and  $h_n(i)$ is the channel power gain between the AP and user $n$ in time slot $i$.

In the following, we propose three optimal multiuser scheduling schemes that control the R-E tradeoff under different fairness considerations.

\section{Optimal Maximum Throughput (MT) Scheme}
First, we consider a scheduling scheme which maximizes the average sum rate subject to a constraint on the minimum required average aggregate harvested energy.  We note that this scheme aims to reveal the best system performance, and  fairness in resource allocation for UTs is not considered. To facilitate the following presentation, we introduce the user selection variables $q_n (i)$, where $i \in \{1, 2, \ldots T\}$ and $n \in \{ 1, \ldots , N \}$. In time slot $i$, if user $n$ is scheduled to perform ID, $q_{n} (i) = 1$, whereas $q_{\bar{n}} (i) = 0, \forall  \bar{n} \neq n$, i.e., all the remaining idle users  harvest energy from the transmitted signal. Now, we formulate the MT optimization problem as follows.
\newpage
\begin{framed}
\begin{Prob}{Maximum Throughput Optimization:}\end{Prob}\vspace*{-3mm}
\begin{equation}
\begin{aligned}
& \underset{q_n (i), \forall n, i}{\mathrm{maximize}}
& & \bar{R}_{\mathrm {sum}} \\
& \text{subject to}
& & \text{C1:} \hspace{7pt} \sum^N_{n=1} q_n (i) = 1, \forall i, \\
&&& \text{C2:} \hspace{7pt} q_n (i) \in \{ 0, 1 \}, \forall n, i, \\
&&& \text{C3:} \hspace{7pt} \bar{Q}_{\mathrm{sum}} \geq Q_{\mathrm{req}},
\end{aligned}
\label{eq:MTbin}
\end{equation}
\end{framed}\vspace*{-8mm}\hspace*{-5mm}
where
\begin{alignat}{2}
\bar{R}_{\mathrm {sum}} &= \lim_{T \to \infty} \frac{1}{T} \sum^{T}_{i=1}\sum^{N}_{n=1} q_n (i) C_n (i), \\
 \bar{Q}_{\mathrm {sum}} &= \lim_{T \to \infty} \frac{1}{T} \sum^{T}_{i=1}\sum^{N}_{n=1} (1-q_n (i)) Q_n (i),\, \mbox{and} \\
 C_n(i) &= \log_2 \left(1 + \frac{P h_n(i)}{\sigma_n^2}\right).
\end{alignat}
Here, $\sigma_n^2$ is the additive white Gaussian noise power at UT $n$. In the considered problem, we focus on the long-term system performance for $T \rightarrow \infty$. Constraints C1 and C2 ensure that in each time slot only one user is selected to receive information. C3 ensures that the average amount of harvested energy $\bar{Q}_{\mathrm {sum}}$ is no less than the minimum required amount $Q_{\mathrm {req}}$. Since the user selection variables $q_n (i), \forall n, i$, are binary, the problem in \eqref{eq:MTbin} is non-convex. In order to handle the non-convexity, we adopt the time-sharing relaxation. In particular, we relax the binary constraint C2 such that $q_n (i)$ is a continuous value between zero and one.  Then, the relaxed version of problem  (\ref{eq:MTbin}) can be written  in minimization form as:\begin{framed}
\begin{equation}
\begin{aligned}
& \underset{q_n (i), \forall n, i}{\mathrm{minimize}}
& & -\bar{R}_{\mathrm {sum}} \\
& \text{subject to}
& & \text{C1, C3}, \\
&&& \widetilde{\text{C2}}: \hspace{7pt}  0 \leq q_n (i) \leq 1, \forall n, i.
\end{aligned}
\label{eq:MTrel}
\end{equation}\end{framed}\vspace*{-8mm}\hspace*{-5mm}
Now,  we introduce the following theorem that reveals the tightness of the binary constraint relaxation.
\begin{Thm}Problems in (\ref{eq:MTbin}) and (\ref{eq:MTrel}) are equivalent\footnote{Here, ``equivalent" means that both problems share the same optimal $q_n(i)$. } with probability one, when $h_n (i), \forall n, i$ are independent and continuously distributed. In particular, the constraint relaxation of C2 is tight, i.e.,\end{Thm}\vspace*{-4mm}
\begin{equation}
\mbox{C2} \Leftrightarrow \widetilde{\text{C2}}: 0 \leq q_n (i) \leq 1, \forall n, i. \label{eq:Theorem}
\end{equation}

\begin{proof}
Theorem 1 will be proved in the following based on the optimal solution of (\ref{eq:MTrel}).
\end{proof}

The constraint relaxed problem is convex with respect to the optimization variables and satisfies the Slater's constraint qualification, therefore strong duality holds and the duality gap is zero. Hence, the optimal solution of the primal problem is equal to the optimal solution of the dual problem. We solve \eqref{eq:MTrel} via the dual problem to get insights into the structure of the solution.

To this end, we first define the Lagrangian function for the above optimization problem which is given by:
\begin{align}\label{eqn:lag}
&  L(q_n(i), \lambda (i), \alpha_n (i), \beta_n(i), \nu ) = - \bar{R}_{\mathrm{sum}} + \sum^T_{i=1} \lambda (i) \left(
\sum^N_{n=1} q_n(i) -1 \right) \notag \\
& + \sum^T_{i=1} \sum^N_{n=1} \alpha_n(i) \left( q_n(i) -1 \right) - \sum^T_{i=1}\sum^N_{n=1} \beta_n(i) q_n(i) + \nu \left( Q_{\mathrm{req}} - \bar Q_{\mathrm{sum}} \right) \notag \\
& = - \frac{1}{T} \sum^T_{i=1} \sum^N_{n=1} q_n (i) C_n (i) + \sum^T_{i=1} \lambda (i) \left( \sum^N_{n=1} q_n(i) -1 \right) + \sum^T_{i=1} \sum^N_{n=1} \alpha_n(i) \left( q_n(i) -1 \right) \notag \\
& - \sum^T_{i=1}\sum^N_{n=1} \beta_n(i) q_n(i) + \nu \left( Q_{\mathrm{req}} - \frac{1}{T} \sum^{T}_{i=1}\sum^{N}_{n=1} (1-q_n (i)) Q_n (i) \right) \notag \\
& = \sum^T_{i=1} \sum^N_{n=1} q_n (i) \left( - \frac{1}{T} C_n (i) + \lambda (i) + \alpha_n (i) - \beta_n (i) + \nu \frac{1}{T} Q_n (i) \right) - \sum_{i=1}^T \lambda (i) \notag\\
& - \sum_{i=1}^T \sum_{n=1}^N \alpha_n(i) + \nu Q_{\mathrm{req}} - \nu \frac{1}{T} \sum_{i=1}^T \sum_{n=1}^N Q_n (i) ,
\end{align}
 where $\lambda (i), \beta_n(i), \alpha_n (i)$ and $\nu$ are the Lagrange multipliers corresponding to constraints C1, C2, C3, and C4, respectively. Thus, the dual problem of (\ref{eq:MTrel}) is given by
\begin{eqnarray}
\underset{\alpha_n (i), \beta_n(i)\ge 0, \lambda (i)}{\mathrm{maximize}} \underset{q_n(i)}{\mino}\,  L(q_n(i), \lambda (i), \alpha_n (i), \beta_n(i), \nu ).\hspace{-5pt}
\end{eqnarray}
In order to determine the optimal user selection policy, we apply standard convex optimization techniques via  the examination of the Karush-Kuhn-Tucker (KKT) conditions which are summarized in the following:
\begin{enumerate}

\vspace{-1pt}
\item Stationarity condition: the differentiation of the Lagrangian function with respect to the primal variables $q_n (i)$ $\forall n, i$ is equal to zero at the optimum point, i.e.,
\begin{equation}
\frac{\partial L}{\partial q_n(i)} = 0, \hspace{10pt} \forall i,n.
\end{equation}

\vspace{-12pt}
\item Primal feasibility condition: the optimal solution has to satisfy the constraints of the primal problem.

\vspace{-8pt}
\item Dual feasibility condition: the Lagrange multipliers for the inequality constraints have to be non-negative, i.e.,
\begin{subequations}\label{eqn:dual_fea_MT}
\begin{alignat}{3}
& \alpha_n(i)  \geq 0, \hspace{10pt} \forall i,n, \\
& \beta_n(i) \geq 0, \hspace{10pt} \forall i,n, \\
& \nu \geq 0.
\end{alignat}
\end{subequations}

\vspace{-12pt}
\item Complementary slackness: if an inequality is inactive, i.e., the optimal solution is in the interior of the corresponding set, the corresponding Lagrange multipliers are zeros, i.e.,
\begin{subequations}
\begin{alignat}{3}
& \alpha_n(i) \left( q_n(i) -1 \right) = 0, & \hspace{10pt} \forall i,n, \label{eqn:slackness_conditions_MT1}  \\
& \beta_n(i) q_n(i) = 0, &\forall i,n, \label{eqn:slackness_conditions_MT2} \\
& \nu \left( Q_{\mathrm{req}} - \bar{Q}_{\mathrm{sum}} \right) = 0. \label{eqn:slackness_conditions_MT3}
\end{alignat}
\end{subequations}
\end{enumerate}

In order to determine the optimal selection policy, we differentiate the Lagrangian in \eqref{eqn:lag} with respect to $q_n (i)$ and set it to zero:
\begin{equation}\label{eqn:partial_d_MT}
\frac{\partial L}{\partial q_n(i)} = - \frac{1}{T} C_n(i) + \lambda(i) + \alpha_n(i) - \beta_n(i) +
\frac{1}{T} \nu Q_n(i) = 0, \hspace{10pt} \forall i,n.
\end{equation}
We define $n^*$ as the index of the user which should be optimally selected in time slot $i$, i.e., $q_{n^*} (i) = 1$. Then, the necessary conditions for $q_{n^*}(i) = 1$ are
\begin{subequations}\label{eqn:optimality_conditions_MT}
\begin{alignat}{3}
& q_n(i) = 0, & \hspace{10pt} \forall n \neq n^*,\label{eqn:optimality_conditions_MT1} \\
& \alpha_n(i) = 0, & \forall n \neq n^*,\label{eqn:optimality_conditions_MT2} \\
& \beta_{n^*}(i) = 0,\label{eqn:optimality_conditions_MT3}
\end{alignat}
\end{subequations}
where \eqref{eqn:optimality_conditions_MT1} follows from C1 in \eqref{eq:MTrel}, \eqref{eqn:optimality_conditions_MT2} follows from \eqref{eqn:slackness_conditions_MT1}, and \eqref{eqn:optimality_conditions_MT2} follows from \eqref{eqn:slackness_conditions_MT1}.

In the following, we substitute \eqref{eqn:optimality_conditions_MT} into \eqref{eqn:partial_d_MT} and introduce the selection metric $\Lambda_n (i)$ as
\begin{subequations}\label{eqn:selection_metric_cal_MT}
\begin{alignat}{3}\label{eqn:selection_metric_cal_MT1}
& \Lambda_{n^*}(i) = T \left( \lambda (i) + \alpha_{n^*}(i) \right) = C_{n^*}(i) - \nu^* Q_{n^*}(i), \\
& \Lambda_n(i) = T \left( \lambda (i) - \beta_n(i) \right) = C_n(i) - \nu^* Q_n(i), \hspace{10pt} \forall n \neq n^*.
\label{eqn:selection_metric_cal_MT2}
\end{alignat}
\end{subequations}

By subtracting \eqref{eqn:selection_metric_cal_MT2} from \eqref{eqn:selection_metric_cal_MT1}, we obtain
\begin{equation}
 \Lambda_{n^*}(i) - \Lambda_n(i) = T \left( \alpha_{n^*}(i) + \beta_n (i) \right).
\end{equation}

From the dual feasibility conditions in \eqref{eqn:dual_fea_MT}, we know that $\alpha_n(i) \geq 0$ and $\beta_n(i) \geq 0$ which yields
\begin{subequations}
\begin{alignat}{3}
& \alpha_{n^*} (i) + \beta_n(i) \geq 0, \\
& \Lambda_{n^*}(i) \geq \Lambda_n(i), \hspace{10pt} \forall n \neq n^*.
\end{alignat}
\end{subequations}

We note that the probability that $\Lambda_{n^*}(i) = \Lambda_n(i)$ is zero $\forall n \neq n^*,\forall i$ since  $\Lambda_n(i)\,\forall n$, are continuous random variables. Thus, the selection criterion for the MT scheme reduces to
\begin{equation}\label{eqn:selection_metric}
\Lambda_{n^*} (i) = \max \limits_{n \in \{ 1, \ldots, N \}} \{ C_n(i) - \nu^* Q_n(i) \},
\end{equation}
where the Lagrange multiplier $\nu^*$ is chosen such that the constraint on the harvested energy C4 is satisfied. $\nu^*$ works as threshold  which depends only on the long-term statistics of the channels. Thus, it can be calculated offline and used for online multiuser scheduling as long as the channel statistics remain unchanged. We note that although the original problem in \eqref{eqn:selection_metric} considers infinite number of time slots and long-term averages for the sum rate and the total harvested energy, interestingly, the optimal scheduling rule in \eqref{eqn:selection_metric} depends only on the current time slot, i.e., online scheduling is optimal. Besides, the solution of the relaxed problem
is itself of the Boolean type since $q_{n}(i)\in\{0,1\}$. Therefore, the adopted binary relaxation is tight.

On the other hand, the optimal value of $\nu^*$ can be obtained iteratively via the gradient method which is described in Algorithm 1, where $m$ is the iteration index and $\Theta$ is an appropriately chosen step.\newpage

\hspace{-10pt}\rule{\textwidth}{1pt}\vspace{-5pt}
\textbf{Algorithm 1} Gradient algorithm for $\nu^*$ \vspace{-5pt}

\vspace{-5pt}
\hspace{-10pt}\rule{\textwidth}{0.5pt}
\vspace{-22pt}
\begin{algorithmic}
\PRINT{the iteration index $m=0$ and dual variable $\nu[0]$}
\REPEAT
\STATE \begin{enumerate}
\item Generate a sufficiently large number of channel realizations, compute the metric in \eqref{eqn:selection_metric} and make a selection of user $n^*$.
\vspace{-10pt}
\item Compute the average harvested energy $\bar{Q}_{\mathrm{sum}}$.
\vspace{-10pt}
\item Update $\nu [m+1] = \max \{ \nu [m] + \nabla_{\nu} \Theta, 0\}$, where $\nabla_{\nu} = Q_{\mathrm{req}} - \bar{Q}_{\mathrm{sum}} $.
\end{enumerate}
\UNTIL{convergence to $\nu^*$.}
\end{algorithmic}
\vspace{-12pt}
\rule{\textwidth}{0.5pt}
\vspace{1pt}

From \eqref{eqn:selection_metric}, we have the following observations:
\vspace{-5pt}
\begin{itemize}
\item If the minimum required harvested energy is not stringent and is always satisfied, e.g. $Q_{\mathrm{req}}=0$, then the problem formulation is equivalent to the conventional maximum sum-rate scheduling scheme. In this case, the second term of the selection metric in \eqref{eqn:selection_metric} is equal to $0$ and therefore $\nu^* = 0$.
\item For a fixed long-term channel statistic, a more stringent minimum required power transfer $Q_{\mathrm{req}}$ always lead to a larger the value of $\nu^*$.
\end{itemize}

\section*{Simulation Results}
Now, we perform simulations for the designed optimal resource allocation scheme.  The important simulation parameters are summarized in Table 1. We adopt the path loss model from \cite{Rappaport} and the UTs are randomly and uniformly distributed between the reference distance and maximum service distance.

\begin{table}\caption{Simulation parameters.}

\begin{center}
\begin{tabular}{ll}
\hline
\textbf{Parameter}    & \textbf{Value} \\
\hline
AP transmit power $P$ & $40$ dBm \\
Noise power $\sigma^2_n$ & $-62$ dBm \\
RF-to-DC conversion efficiency $\xi_n$ & $0.5$ \\
Path loss exponent & $3.6$ \\
Maximum service distance &  $100$ m \\
Reference distance & $2$ m \\
Antenna gain of AP and UTs & $10$ dBi \& $2$ dBi \\
Carrier center frequency & $915$ MHz \\
Bandwidth & $200$ kHz \\
Fading channel & Rayleigh\\
\hline
\end{tabular}
\end{center}
\label{tab:res}
\end{table}

We use the order-based SNR scheduler from \cite{Morsi2014} as a baseline scheme. This scheduler performs user selection according to the following rule:
\begin{equation}
n^* (i) = \argorder\limits_{n\in\{1,\ldots,N\}} h_n (i)
\end{equation}
where $\argorder$ is defined as the argument of a certain selection order $j \in \{1, \ldots, N \}$. That is, the user whose channel power gain $h_n(i)$ has order $j$ is scheduled for ID.

\begin{figure}[!h]
\centering
\includegraphics[width= 4.5 in]{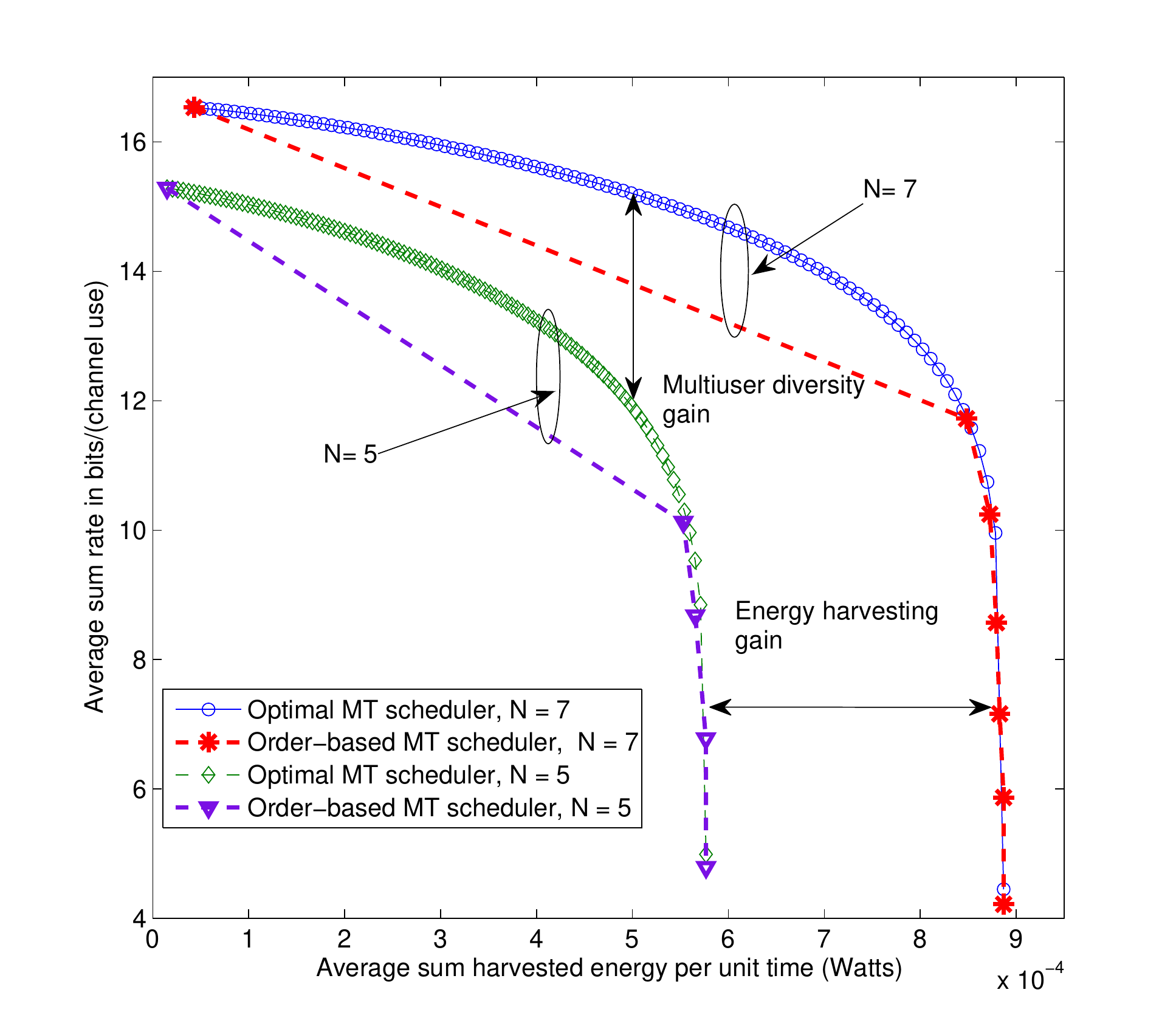}
\caption{Average sum rate versus average sum harvested energy of the MT schemes for different numbers of UTs.}
\label{fig:MT2}
\end{figure}

Figure \ref{fig:MT2} shows the average sum rate (bits/(channel use)) versus the average sum harvested energy (Watts) of the MT schemes for different numbers of users. We note that the suboptimal order-based scheme can only achieve discrete points on the R-E curves, corresponding to the selection orders $j\in\{1,\ldots,N\}$.   On the contrary, the proposed optimal MT scheduling scheme can achieve any feasible point on the R-E curve, which provides a higher flexibility for the system designer to strike a balance between average sum rate and average harvested energy. Besides, as expected,  the average system sum rate  increases with the number of UTs as the proposed scheme is able to exploit multiuser diversity. Furthermore, the average sum harvested energy also increases with the number of UTs since more idle users participate in EH in any given time slot.

\newpage
\section{Optimum Proportional Fair (PF) Scheme}
In the MT scheme, UTs with weak channel conditions may be deprived from gaining access to the channel which leads to user starvation. In order to strike a balance between system throughput and fairness, we introduce proportional fairness  into our scheduler, which aims to provide each UT with a performance proportional to its channel conditions. This is achieved by allowing all UTs to access the channel with  equal chances. In this case, the optimization problem with the relaxed binary constraint on the user selection variables is formulated as:\vspace*{-4mm}
\begin{framed}\vspace*{-3mm}
\begin{Prob}{Optimal Proportional Fair Optimization:}\end{Prob}\vspace*{-4mm}
\begin{equation}
\begin{aligned}
& \underset{q_n (i), \forall i, n}{\mathrm{maximize}}
& & \bar{R}_{\mathrm{sum}} \\
& \text{subject to}
& & \text{C1:} \hspace{7pt}  \sum^N_{n=1} q_n (i) = 1, \forall i, \\
&&& \text{C2:} \hspace{7pt}  q_n (i) \left( 1 - q_n (i) \right) = 0, \forall n, i, \\
&&& \text{C3:} \hspace{7pt}  \bar{Q}_{\mathrm{sum}} \geq Q_{\mathrm{req}}, \\
&&& \text{C4:} \hspace{7pt}  \sum^T_{i=1} q_n (i) = \frac{T}{N},
\end{aligned}
\end{equation}\vspace*{-4mm}\end{framed}\vspace*{-2mm}\hspace*{-5mm}
where constraint C4 specifies that each user has to access the channel for $\frac{T}{N}$ number of time slots, $T$ is the total number of time slots, and $N$ is the number of users in the system.

 Since $q_n (i), \forall n, i$, is non-convex due to its binary nature, we relax the binary constraint to obtain a convex problem. We note that the adopted constraint relaxation is tight and the proof is given in Appendix A.1. Thus, we obtain the following reformulated problem in minimization form:\vspace*{-4mm}
\begin{framed}\vspace*{-4mm}\begin{equation}\label{eqn:min_problem}
\begin{aligned}
& \underset{q_n (i), \forall i, n}{\text{minimize}}
& & -\bar{R}_{\mathrm{sum}} \\
& \text{subject to}
& & \text{C1:} \hspace{7pt}  \sum^N_{n=1} q_n (i) - 1 = 0, \forall i, \\
&&& \text{C2:} \hspace{7pt}  -q_n (i) \leq 0, \forall n, i,  \\
&&& \text{C3:} \hspace{7pt}  q_n (i) -1 \leq 0, \forall n, i, \\
&&& \text{C4:} \hspace{7pt}  Q_{\mathrm{req}} - \bar{Q}_{\mathrm{sum}} \leq 0,  \\
&&& \text{C5:} \hspace{7pt}  \frac{1}{T} \sum^T_{i=1} q_n (i) - \frac{1}{N} = 0.
\end{aligned}
\end{equation}
\end{framed}
The Lagrangian function for the problem \eqref{eqn:min_problem} is given by:
\begin{align}
& L(q_n(i), \lambda (i), \alpha_n (i), \beta_n(i), \nu, \gamma_n ) = - \bar{R}_{\mathrm{sum}} + \sum^T_{i=1} \lambda (i) \left( \sum^N_{n=1} q_n(i) -1 \right) \notag \\
& + \sum^T_{i=1} \sum^N_{n=1} \alpha_n(i) \left( q_n(i) -1 \right)- \sum^T_{i=1}\sum^N_{n=1} \beta_n(i) q_n(i) + \nu \left( Q_{\mathrm{req}} - \bar Q_{\mathrm{sum}} \right) + \sum^N_{n=1} \gamma_n
\left( \frac{1}{T} \sum^T_{i=1} q_n (i) - \frac{1}{N} \right) \notag \\
& = - \frac{1}{T} \sum^T_{i=1} \sum^N_{n=1} q_n (i) C_n (i) + \sum^T_{i=1} \lambda (i) \left( \sum^N_{n=1} q_n(i) -1 \right) + \sum^T_{i=1} \sum^N_{n=1} \alpha_n(i) \left( q_n(i) -1 \right) \notag \\
& - \sum^T_{i=1}\sum^N_{n=1} \beta_n(i) q_n(i) + \nu \left( Q_{\mathrm{req}} - \frac{1}{T} \sum^{T}_{i=1}\sum^{N}_{n=1} (1-q_n (i)) Q_n (i) \right) + \sum^N_{n=1} \gamma_n
\left( \frac{1}{T} \sum^T_{i=1} q_n (i) - \frac{1}{N} \right) \notag \\
& = \sum^T_{i=1} \sum^N_{n=1} q_n (i) \left( - \frac{1}{T} C_n (i) + \lambda (i) + \alpha_n (i) - \beta_n (i) + \nu \frac{1}{T} Q_n (i) + \frac{1}{T} \gamma_n \right) - \sum_{i=1}^T \lambda (i) \notag \\
& - \sum_{i=1}^T \sum_{n=1}^N \alpha_n(i) + \nu Q_{\mathrm{req}} - \nu \frac{1}{T} \sum_{i=1}^T \sum_{n=1}^N Q_n (i) - \sum_{n=1}^N \gamma_n \frac{1}{N},
\end{align}
where $\lambda (i), \beta_n(i), \alpha_n (i), \nu$, and $\gamma_n$ are the Lagrange multipliers corresponding to constraints C1, C2, C3, C4, and C5, respectively.

We study the structure of the optimal scheduling policy via the KKT conditions. The stationarity condition is then given by
\begin{equation}
\frac{\partial L}{\partial q_n(i)} = - \frac{1}{T} C_n(i) + \lambda(i) + \alpha_n(i) - \beta_n(i) +
\frac{1}{T} \nu Q_n(i) + \frac{1}{T} \gamma_n = 0.
\end{equation}

Analogously to the MT scheme, we introduce the selection metric $\Lambda_n (i)$ as
\begin{subequations}\label{eqn:selection_metric_cal_PF}
\begin{alignat}{3}
& \Lambda_{n^*}(i) = T \left( \lambda (i) + \alpha_{n^*}(i) \right) = C_{n^*}(i) - \nu^* Q_{n^*}(i) - \gamma^*_{n^*}, \label{eqn:selection_metric_cal_PF1} \\
& \Lambda_n(i) = T \left( \lambda (i) - \beta_n(i) \right) = C_n(i) - \nu^* Q_n(i) - \gamma^*_n, \hspace{10pt} \forall n \neq n^*,\label{eqn:selection_metric_cal_PF2}
\end{alignat}
\end{subequations}
where $n^*$ is the user that is optimally scheduled for information reception. Subtracting \eqref{eqn:selection_metric_cal_PF2} from \eqref{eqn:selection_metric_cal_PF1} we obtain
\begin{equation}
 \Lambda_{n^*}(i) - \Lambda_n(i) =T \left(  \alpha_{n^*}(i) + \beta_n (i) \right).
\end{equation}

From the dual feasibility conditions, it follows that $\alpha_n(i) \geq 0$, $\beta_n(i) \geq 0$, therefore
\begin{subequations}
\begin{alignat}{3}
& \alpha_{n^*} (i) + \beta_n(i) \geq 0, \\
& \Lambda_{n^*}(i) \geq \Lambda_n(i), \hspace{10pt} \forall n \neq n^*.
\end{alignat}
\end{subequations}

We note that the probability that $\Lambda_{n^*}(i) = \Lambda_n(i)$ is zero, since $\Lambda_n(i)$ are continuous random variables. Hence, the selection criterion for the PF scheme reduces to
\begin{equation}\label{eqn:selection_metric_PF}
\Lambda_{n^*} (i) = \max \limits_{n \in \{ 1, \ldots, N \}} \{ C_n (i) - \nu^* Q_n (i) - \gamma^*_n \},
\end{equation}
where Lagrange multiplier $\nu^*$ for constraint C4 ensures that at least $Q_{\mathrm{req}}$  amount of energy is harvested.  Lagrange multipliers $\gamma^*_n$  for constraint C3 guarantee that each user accesses the channel equal number of times. Analogously to the MT scheme, $\nu^*$ and $\gamma^*_n$ depend  only on the long-term statistics of the channels,  therefore they can be calculated offline and used for online multiuser scheduling as long as the channel statistics remain unchanged. Besides, the optimal scheduling interestingly depends only on the current time slot, i.e., online scheduling is optimal. We note that the optimal PF scheduling rule  is similar to the MT scheduling rule in \eqref{eqn:selection_metric_PF}, but the PF selection metric in (\ref{eqn:UT_metric}) contains an additional term $\gamma^*_n$ that provides proportional fairness.

The optimal values of $\nu^*$ and $\gamma_n^*$ can be obtained iteratively via the gradient method as described in Algorithm 2, where $m$ is the iteration index. Variables $\Theta$ and $\Gamma$ are appropriately chosen step sizes to facilitate the convergence of the gradient method.\vspace{5pt}

\hspace{-10pt}\rule{\textwidth}{1pt}\vspace{-5pt}
\textbf{Algorithm 2} Gradient algorithm for $\nu^*$ and $\gamma_n^*$ \vspace{-5pt}

\vspace{-5pt}
\hspace{-10pt}\rule{\textwidth}{0.5pt}
\vspace{-22pt}
\begin{algorithmic}
\PRINT{the iteration index $m=0$, $\nu[0]$, and $\gamma_n [0]$}
\REPEAT
\STATE \begin{enumerate}
\item Generate a sufficiently large number of channel realizations, compute the metric in \eqref{eqn:selection_metric_PF} and make a selection of user $n^*$.
\vspace{-10pt}
\item Compute the average harvested energy $\bar{Q}_{\mathrm{sum}}$ and the number of selections of each user.
\vspace{-10pt}
\item Update $\nu [m+1] = \max \{ \nu [m] + \nabla_{\nu} \Theta, 0\}$, where $\nabla_{\nu} = Q_{\mathrm{req}} - \bar{Q}_{\mathrm{sum}} $,

update $\gamma_n [m+1] = \gamma_n [m] + \nabla_{\gamma} \Gamma $, where $\nabla_{\gamma} = \frac{1}{T} \sum^T_{i=1} q_n (i) - \frac{1}{N}$.
\end{enumerate}
\UNTIL{convergence to $\nu^*$ and $\gamma_n^*$.}
\end{algorithmic}
\vspace{-12pt}
\rule{\textwidth}{0.5pt}
\vspace{1pt}

\section*{Simulation Results}
Next, we perform simulations for the scheme investigated in this section. The simulation parameters are assumed the same as in Section 2.2. As for the baseline scheme, we will use the order-based normalized-SNR (N-SNR) scheduler from \cite{Morsi2014}, which performs user selection according to the following rule:
\begin{equation}
n^* (i) = \argorder\limits_{n\in\{1,\ldots,N\}} \frac{h_n (i)}{\Omega_n},
\end{equation}
where $\Omega_n$ denotes the mean channel power gain of UT $n$.  This scheduling rule also ensures PFness, i.e., all users gain access to the channel with equal number of times.

\begin{figure}[t]  \centering
\includegraphics[width=4.5 in]{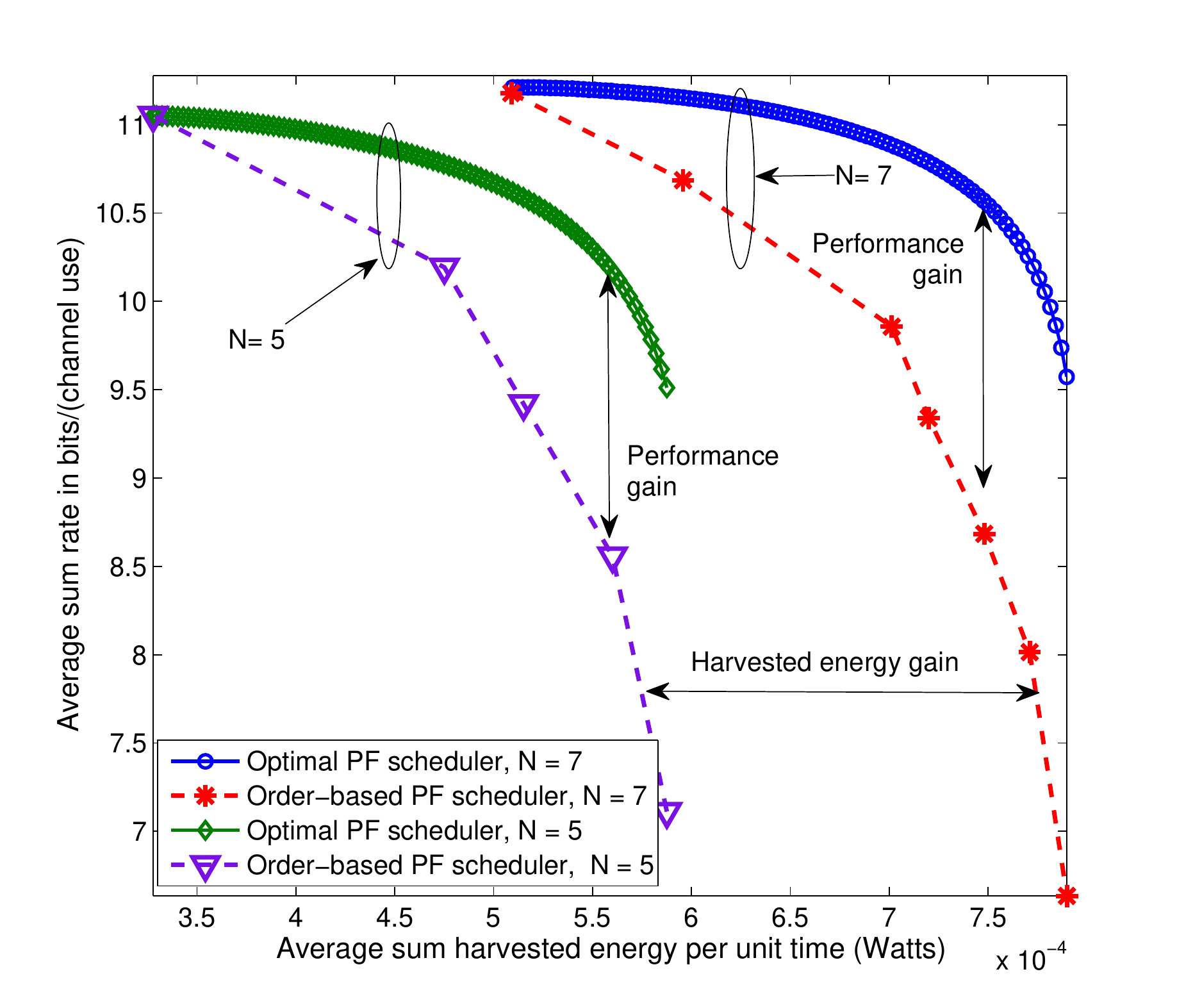}
\caption{Average sum rate versus average sum harvested energy of the PF schemes for different numbers of UTs.}
\label{fig:PropFairRice}
\end{figure}

Figure \ref{fig:PropFairRice}   depicts the average sum rate (bits/(channel use)) versus the average sum harvested energy (Watts) for the PF. It can be seen that the feasible R-E region of all schemes decreases compared to the MT scheduler in Figure \ref{fig:MT2}. This is because the PF scheduler takes fairness into account in the resource allocation and, as a result, cannot fully exploit the multiuser diversity for improving the average system sum rate. On the other hand, it can be seen that our proposed optimal schemes provide a substantial average sum rate  gain compared to the corresponding suboptimal order-based scheme, especially for a high amount of average harvested energy in the system.  In fact, the proposed optimization framework provides more degrees of freedom across different time slots in resource allocation compared to the suboptimal scheduling schemes.  This allows the system to exploit the multiuser diversity to some extent for resource allocation even if fairness is taken into consideration.\vspace*{-5mm}
\section{Optimum Equal Throughput (ET) Scheme}
Although the PF scheduler enables equal channel access probability for all UTs, it does not provide any guaranteed minimum data rate to them.  On the contrary, the ET criterion is more fair from the users' prospective compared to the PF criterion, as all the UTs achieve the same average throughput asymptotically for $T\rightarrow\infty$. Therefore, in this section, we design a scheduler which achieves ET fairness.  Thus, the objective is to maximize the minimum average achievable rates among all the UTs, i.e., maximize $\min\limits_{n} \bar{C}_n$ where $\bar{C}_n=\lim_{T \to \infty} \frac{1}{T} \sum^{T}_{i=1} q_n (i) C_n (i)$. The optimization problem is then formulated as follows:\begin{framed}\vspace*{-3mm}
\begin{Prob}{Optimal Equal Throughput Optimization:}\end{Prob}\vspace*{-6mm}
\begin{equation}\label{eqn:max_min_ET}
\begin{aligned}
& \underset{q_n (i), \forall i, n}{\mathrm{maximize}}
& & \min \bar{C}_n \\
& \text{subject to}
& & \text{C1:} \hspace{7pt}  \sum^N_{n=1} q_n (i) = 1, \forall i, \\
&&& \text{C2:} \hspace{7pt}  q_n (i) \left( 1 - q_n (i) \right) = 0, \forall n, i, \\
&&& \text{C3:} \hspace{7pt}  \bar{Q}_{\mathrm{sum}} \geq Q_{\mathrm{req}}.
\end{aligned}
\end{equation}\vspace*{-3mm}\end{framed}\vspace*{-4mm}
Equation \eqref{eqn:max_min_ET} is a max-min optimization problem, which can be rewritten in its equivalent hypograph form:
\begin{framed}\vspace*{-4mm}\begin{equation}\label{eqn:max_min_ET_hypo}
\begin{aligned}
& \underset{r, q_n (i), \forall i, n}{\mathrm{maximize}}
& & r \\
& \text{subject to}
& & \text{C1:} \hspace{7pt}  \bar{C}_n \geq r, \hspace{5pt} \forall n, \\
&&& \text{C2:} \hspace{7pt}  \sum^N_{n=1} q_n (i) = 1, \forall i, \\
&&& \text{C3:} \hspace{7pt}  q_n (i) \left( 1 - q_n (i) \right) = 0, \forall n, i, \\
&&& \text{C4:} \hspace{7pt}  \bar{Q}_{\mathrm{sum}} \geq Q_{\mathrm{req}}.
\end{aligned}
\end{equation}\end{framed}\vspace*{-3mm}\hspace*{-5mm}
$r$ is an auxiliary optimization variable for handling the max-min objective function. After binary relaxation of $q_n (i), \forall n, i$, we formulate our convex optimization problem in minimization form as follows
\vspace{-10pt}\begin{framed}\vspace*{-3mm}\begin{equation}\label{eqn:max_min_ET_hypo_relaxed}
\begin{aligned}
& \underset{r, q_n (i), \forall i, n}{\text{minimize}}
& & - r \\
& \text{subject to}
& & \text{C1, C2, C5,}  \\
&&& \text{C3:} \hspace{7pt}  -q_n (i) \leq 0, \hspace{5pt} \forall n, i,  \\
&&& \text{C4:} \hspace{7pt}  q_n (i) -1 \leq 0, \hspace{5pt} \forall n, i, \\
\end{aligned}
\end{equation}\vspace*{-6mm}\end{framed}\hspace*{-5mm}
We note that the adopted binary constraint relaxation is tight and please refer to Appendix A.1 for the proof.

The Lagrangian function for the problem \eqref{eqn:max_min_ET_hypo_relaxed} is then given by
\begin{align}
& L(q_n(i), \lambda (i), \alpha_n (i), \beta_n(i), \nu, \theta_n ) = -r + \sum^T_{i=1} \lambda (i) \left( \sum^N_{n=1} q_n(i) -1 \right) + \sum^T_{i=1} \sum^N_{n=1} \alpha_n(i) \left( q_n(i) -1 \right) \notag \\
& - \sum^T_{i=1}\sum^N_{n=1} \beta_n(i) q_n(i) + \nu \left( Q_{\mathrm{req}} - \bar Q_{\mathrm{sum}} \right) + \sum^N_{n=1} \theta_n
\left( r- \bar{C}_n \right) \notag \\
& = - r + \sum^T_{i=1} \lambda(i) \sum^N_{n=1} q_n(i) - \sum^T_{i=1} \lambda (i) + \sum^T_{i=1} \sum^N_{n=1} \alpha_n(i) q_n(i) - \sum^T_{i=1} \sum^N_{n=1} \alpha_n(i) - \sum^T_{i=1}\sum^N_{n=1} \beta_n(i) q_n(i) \notag  \\
& + \nu Q_{\mathrm{req}} - \nu \frac{1}{T} \sum^{T}_{i=1}\sum^{N}_{n=1}Q_n(i) + \nu \frac{1}{T} \sum^{T}_{i=1}\sum^{N}_{n=1} q_n (i) Q_n (i) + \sum^N_{n=1} \theta_n r - \sum^N_{n=1} \theta_n \frac{1}{T} \sum^T_{i=1} C_n (i) q_n(i)  \notag \\
& = -r + \sum^T_{i=1} \sum^N_{n=1} q_n (i) \left( \lambda (i) + \alpha_n (i) - \beta_n (i) + \nu \frac{1}{T} Q_n (i) - \theta_n \frac{1}{T} C_n (i) \right) \notag \\
&- \sum_{i=1}^T \lambda (i) - \sum_{i=1}^T \sum_{n=1}^N \alpha_n(i) + \nu Q_{\mathrm{req}} - \nu \frac{1}{T} \sum_{i=1}^T \sum_{n=1}^N Q_n (i) + \sum_{n=1}^N \theta_n r,
\end{align}
where $\theta_n, \lambda (i), \beta_n(i), \alpha_n (i)$, and $\nu$ are the Lagrange multipliers corresponding to constraints C1, C2, C3, C4, and C5, respectively.

We study the structure of the optimal scheduling policy via the KKT conditions. The stationarity condition is expressed as
\begin{subequations}
\begin{alignat}{3}
& \frac{\partial L}{\partial q_n(i)} = \lambda(i) + \alpha_n(i) - \beta_n(i) +
\nu \frac{1}{T} Q_n(i) - \theta_n \frac{1}{T} C_n (i) = 0, \\
& \frac{\partial L}{\partial r} = -1 + \sum^N_{n=1} \theta_n = 0 \Longrightarrow \sum^N_{n=1} \theta_n = 1.
\label{eqn:gradient_ET2}
\end{alignat}
\end{subequations}

Next, we introduce the selection metric $\Lambda_n (i)$ for our ET scheduler as
\begin{subequations}
\begin{alignat}{3}\label{eqn:ET_metric1}
& \Lambda_{n^*}(i) = T \left( \lambda (i) + \alpha_{n^*}(i) \right) = \theta^*_{n^*} C_{n^*}(i) - \nu^* Q_{n^*}(i), \\
& \Lambda_n(i) = T \left( \lambda (i) - \beta_n(i) \right) = \theta^*_n C_n(i) - \nu^* Q_n(i), \hspace{10pt} \forall n \neq n^*,
\label{eqn:ET_metric2}
\end{alignat}
\end{subequations}
where $n^*$ is the optimally scheduled user index.

Subtracting  \eqref{eqn:ET_metric2} from  \eqref{eqn:ET_metric1}, we get
\begin{equation}
\Lambda_{n^*}(i) - \Lambda_n(i) =T \left(  \alpha_{n^*}(i) + \beta_n (i) \right).
\end{equation}

From the dual feasibility conditions, we know that $\alpha_n(i) \geq 0, \beta_n(i) \geq 0$, therefore:
\begin{subequations}
\begin{alignat}{3}
& \alpha_{n^*} + \beta_n(i) \geq 0, \\
& \Lambda_{n^*}(i) \geq \Lambda_n(i), \hspace{10pt} \forall n \neq n^*.
\end{alignat}
\end{subequations}

Similar to the MT and PF schemes, $\Lambda_n(i) \hspace{5pt} \forall n $ are continuous random variables. Hence, the probability that $\Lambda_{n_1}(i) = \Lambda_{n_2}(i)$ for $n_1 \neq n_2$ is zero.
Thus, we obtain the following selection criterion for the ET scheme:
\begin{equation}\label{eqn:selection_metric_ET}
\Lambda_{n^*} (i) = \max \limits_{n \in \{ 1, \ldots, N \}} \{ \theta^*_n C_n(i) - \nu^* Q_n(i) \},
\end{equation}
where  Lagrange multiplier $\nu^*$ ensures that constraint C5 for the minimum requirement of harvested energy is satisfied and  Lagrange multipliers $\theta^*_n$ ensure that all the users have ET. Analogously to the previously described schemes, $\nu^*$ and $\theta^*_n$ only depend on the long-term statistics of the channels, therefore they can be calculated offline and used for online multiuser scheduling as long as the channel statistics remain unchanged.

The optimal values of $\nu^*$ and $\theta_n^*$ can be obtained iteratively via the gradient method described in Algorithm 3, where $m$ is the iteration index and the appropriately chosen step sizes $\Theta$ and $\zeta$ guarantee convergence of $\nu$ and $\theta_n$  to the optimal dual variables $\nu^*$ and $\theta_n^*$, respectively. We note that $\theta_n^* \in [ 0,1 ]$. This is because $\theta_n \geq 0$ from the dual feasibility condition of the inequality constraint C1 and $\theta_n \leq 1$ since $\sum^N_{n=1} \theta_n = 1$ from  \eqref{eqn:gradient_ET2}.

\
\hspace{-10pt}\rule{\textwidth}{1pt}\vspace{-5pt}
\textbf{Algorithm 3} Gradient algorithm for $\nu^*$ and $\theta_n^*$ \vspace{-5pt}

\vspace{-5pt}
\hspace{-10pt}\rule{\textwidth}{0.5pt}
\vspace{-22pt}
\begin{algorithmic}
\PRINT{$m=0$, $\nu[0]$ and $\theta_n [0]$}
\REPEAT
\STATE \begin{enumerate}
\item Generate a sufficiently large number of channel realizations, compute the metric in \eqref{eqn:selection_metric_ET} and make a selection of user $n^*$.
\vspace{-10pt}
\item Compute the average total harvested energy $\bar{Q}_{\mathrm{sum}}$ and the average rate per user $\bar{C}_{n} = \frac{1}{T} \sum^T_{i=1} C_n (i) q_n (i)$.
\vspace{-10pt}
\item Update $\nu [m+1] = \max \{ \nu [m] + \nabla_{\nu} \Theta, 0\}$, where $\nabla_{\nu} = Q_{\mathrm{req}} - \bar{Q}_{\mathrm{sum}} $,

update $\theta_n [m+1] = \left[ \theta_n [m] + \nabla_{\theta} \zeta \right] ^1_0$, where $\nabla_{\theta} = r - \bar{C}_{n}$.
\end{enumerate}
\UNTIL{convergence to $\nu^*$ and $\theta_n^*$}.
\end{algorithmic}
\vspace{-12pt}
\rule{\textwidth}{0.5pt}
\vspace{1pt}

Next, we perform simulations for the proposed optimal ET scheduler. The simulation parameters are assumed to be the same as in Sections 2.2 and 2.3. As for the baseline scheme, we will use the order-based ET scheduler from \cite{Morsi2014}. For this scheme, the users' instantaneous N-SNRs are sorted in ascending ordered, and then among the set of users whose N-SNR orders fall into a predefined set of allowed orders $\Sa$, the AP schedules the one with the minimum moving average throughput. Therefore, the selection rule of the order-based ET scheme is
\begin{equation}
n^* (i) =\argmin\limits_{{O_{\Un_n}\in \Sa}} r_n(i-1),
\end{equation}
where $O_{\Un_n}\in\{1,\ldots,N\}$ is defined as the order of the  instantaneous N-SNR of user $n$, and $r_n(i-1)$ is the throughput of user $n$ averaged over previous time slots up to slot $i-1$.

\begin{figure}[t]
  \centering
\includegraphics[width=4.5 in]{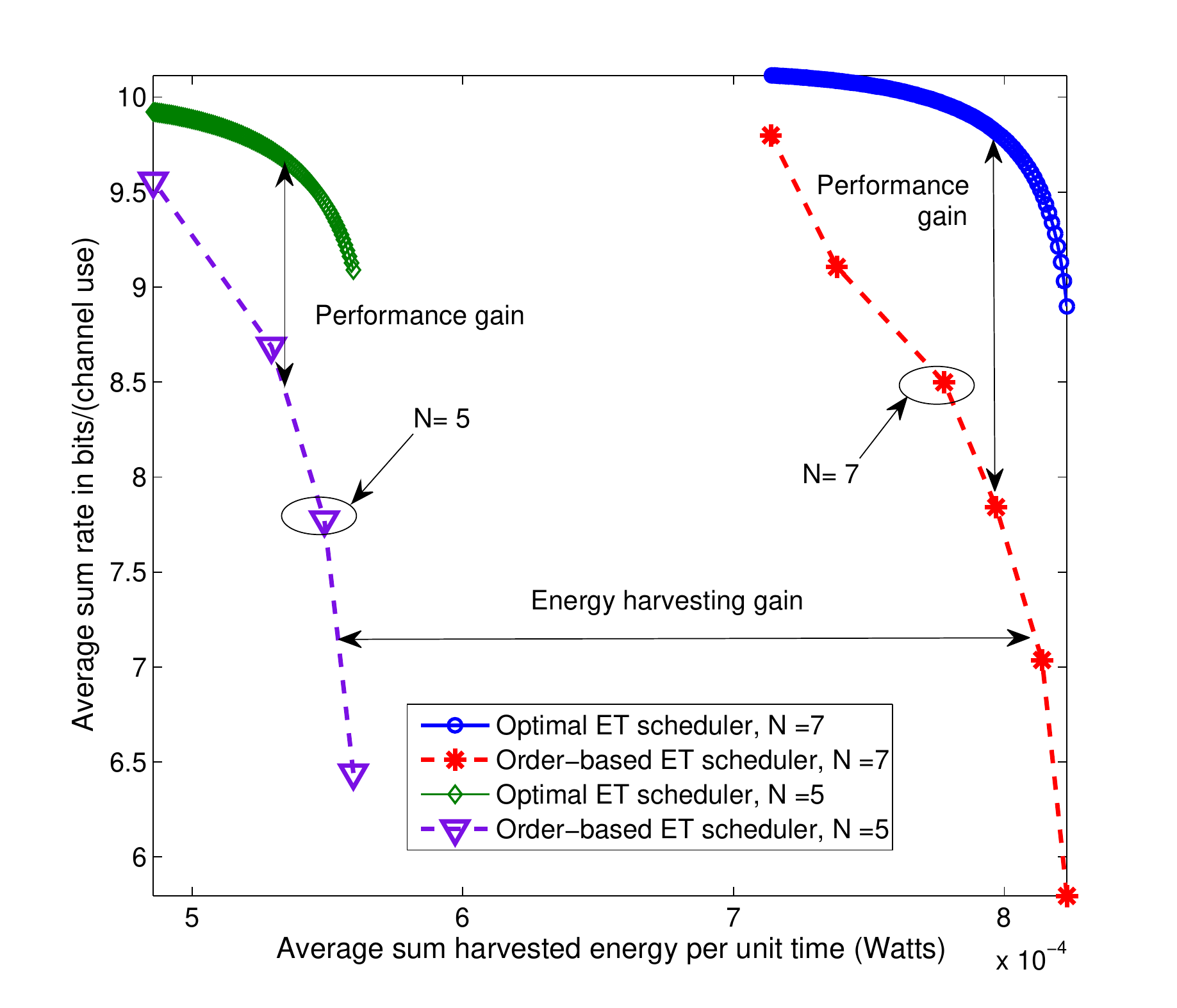}
\caption{Average sum rate versus average sum harvested energy of the ET schemes for different numbers of UTs.}
\label{fig:EqualThRice}
\end{figure}
The R-E curves of the considered schemes are shown in Figure \ref{fig:EqualThRice}. The optimum  ET scheduler provides a substantial sum rate gain compared to the order-based ET scheduler. In particular, the proposed optimal scheduler is able to exploit the degrees of freedom across different time slots for improving the system performance. On the other hand, the average sum harvested energy of the proposed scheme increases rapidly with the numbers of receives in the system due to more UTs participate in EH.  Besides, the average sum rate increases slightly with the numbers of UTs compared to the MT scheduler. In fact, the optimal ET scheme cannot fully exploit the multiuser diversity for improving the average system sum rate due to the required fairness in equal average data rate for all UTs.
\let\cleardoublepage\clearpage
\chapter{Optimum Online Multiuser Scheduling with Power Allocation}
In this chapter, we enhance the MT scheduling scheme (c.f. Section 2.2) by considering joint user scheduling and  power allocation. To this end, we adopt the problem formulation in \eqref{eq:MTbin} and introduce an extra power allocation variable $P_n (i)$ to user $n$ in time slot $i$, as an additional degree of freedom. Thus, we obtain the following problem formulation:\vspace*{-2mm}
\begin{framed}\begin{Prob}{Joint Power and Scheduling Optimization:}\end{Prob}\vspace*{-10mm}
\begin{equation}\label{eqn:power_allocation_formulation}
\begin{aligned}
& \underset{q_n (i), P_n (i), \forall i, n}{\mathrm{maximize}}
& & \frac{1}{T} \sum^{T}_{i=1}\sum^{N}_{n=1} q_n (i) \log_2 \left(1 + \frac{P_n(i) h_n(i)}{\sigma^2}\right) \\
& \text{subject to}
& & \text{C1:} \hspace{7pt} \sum^N_{n=1} q_n (i) = 1, \forall i, \\
&&& \text{C2:} \hspace{7pt} q_n (i) \left( 1 - q_n (i) \right) = 0, \forall n, i, \\
&&& \text{C3:} \hspace{7pt} \sum^N_{n=1} P_n (i) q_n (i) \leq P_{\max}, \forall i, \\
&&& \text{C4:} \hspace{7pt} \frac{1}{T} \sum^{T}_{i=1}\sum^{N}_{n=1} P_n (i) q_n (i) \leq P_{\mathrm{ave}}, \\
&&& \text{C5:} \hspace{7pt} \frac{1}{T} \sum^{T}_{i=1}\sum^{N}_{n=1} (1-q_n(i)) \left( \sum^N_{k=1} P_k (i) q_k (i) \right) \xi_n h_n (i) \geq Q_{\mathrm{req}}.
\end{aligned}
\end{equation}\end{framed}\hspace*{-5mm}
Constraint C3 specifies a hardware constraint which limits the maximum instantaneous transmit power to $P_{\max}$. Constraint C4 constrains the average transmit power budget to $P_{\mathrm{ave}}$. The term $\sum^N_{k=1} P_k (i) q_k (i)$  in C5 represents the total radiated power in time slot $i$.

The objective function in \eqref{eqn:power_allocation_formulation} is non-convex. In order to convexify the objective function, we use the following change of variables: $P'_n (i) = P_n (i) q_n(i) $. Also, we relax the binary constraint on the user selection variables $q_n (i)$ and rewrite \eqref{eqn:power_allocation_formulation} as
\begin{framed}\vspace*{-8mm}
\begin{equation}\label{eqn:power_allocation_formulation2}
\begin{aligned}
& \underset{q_n (i), P'_n (i), \forall i, n}{\mathrm{maximize}}
& & \frac{1}{T} \sum^{T}_{i=1}\sum^{N}_{n=1} q_n (i) \log_2 \left(1 + \frac{P'_n (i) h_n(i)}{q_n (i) \sigma^2}\right) \\
& \text{subject to}
& & \text{C1-C2,}\\
&&& \text{C3:} \hspace{7pt} \sum^N_{n=1} P'_n (i) \leq P_{\max}, \forall i, \\
&&& \text{C4:} \hspace{7pt} \frac{1}{T} \sum^{T}_{i=1}\sum^{N}_{n=1} P'_n (i) \leq P_{\mathrm{ave}}, \\
&&& \text{C5:} \hspace{7pt} \frac{1}{T} \sum^{T}_{i=1}\sum^{N}_{n=1} (1-q_n(i)) \left( \sum^N_{k=1} P'_k (i) \right) \xi_n h_n (i) \geq Q_{\mathrm{req}}.
\end{aligned}
\end{equation}\vspace*{-2mm}\end{framed}
Now the objective function in \eqref{eqn:power_allocation_formulation2} is concave, since $f(x) = \log (1+x)$ is concave and $f(x,y) = y \log (1+ \frac{x}{y})$ is jointly concave with respect to $x$ and $y$ \cite{Boyd2004}. Nevertheless, constraint C6 is still non-convex due to the coupling of the optimization variables. In the following, we adopt the big-M formulation to linearize  the coupled terms $(1-q_n(i)) \left( \sum^N_{k=1} P'_k (i) \right)$. The new problem formulation is given by
\vspace*{-2mm}
\begin{framed}\vspace*{-1mm}
\begin{Prob}{Problem Reformulation:}\end{Prob}\vspace*{-9mm}
\begin{equation}\label{eqn:power_allocation_formulation3}
\begin{aligned}
& \underset{q_n (i), P'_n (i), P^{\mathrm{virtual}}_n (i), \forall i, n}{\mathrm{maximize}}
& & \frac{1}{T} \sum^{T}_{i=1}\sum^{N}_{n=1} q_n (i) \log_2 \left(1 + \frac{P'_n (i) h_n(i)}{q_n (i) \sigma^2}\right) \\
& \text{subject to}
& & \text{C1-C2,}\\
&&& \text{C3:} \hspace{7pt} \sum^N_{n=1} P'_n (i) \leq P_{\max}, \forall i, \\
&&& \text{C4:} \hspace{7pt} \frac{1}{T} \sum^{T}_{i=1}\sum^{N}_{n=1} P'_n (i) \leq P_{\mathrm{ave}}, \\
&&& \text{C5:} \hspace{7pt} \frac{1}{T} \sum^{T}_{i=1}\sum^{N}_{n=1}\left( \sum^N_{k=1} P^{\mathrm{virtual}}_k (i) \right) \xi_n h_n (i) \geq Q_{\mathrm{req}},\\
&&& \text{C6:} \hspace{7pt}P^{\mathrm{virtual}}_n (i)\leq (1-q_n (i)) P_{\max}, \forall n, i,\\
&&& \text{C7:} \hspace{7pt}P^{\mathrm{virtual}}_n (i)\leq  P'_n (i),  \forall n, i,\\
&&& \text{C8:} \hspace{7pt}P^{\mathrm{virtual}}_n (i)\geq 0, \forall n, i, \\
\end{aligned}
\end{equation}\vspace*{-2mm}\end{framed}\hspace*{-5mm}
where $ P^{\mathrm{virtual}}_n (i)$ is auxiliary variables  for solving the problem.  In fact, variable $P^{\mathrm{virtual}}_n (i)$ can be treated as  the virtual transmit power variable which is controlled by $q_{n}(i)$ in constraint C7 and the actual transmit power $P'_n (i)$ in constraint C8. We note that both Problems \eqref{eqn:power_allocation_formulation3} and \eqref{eqn:power_allocation_formulation2} are equivalent when $q_n (i)$ is binary. In particular, both problems share the same optimal solution.  Besides, by following a similar approach as in Appendix A.1., it can be shown that the binary relaxation on $q_n(i)$ is tight at the optimal solution, i.e., $q_n^*(i)\in\{0,1\}$. More importantly, Problem \eqref{eqn:power_allocation_formulation3} is a convex optimization problem which can be solved efficiently via standard numerical solvers designed for convex programs.
\begin{figure}[t]
\centering
\includegraphics[width=4.5 in]{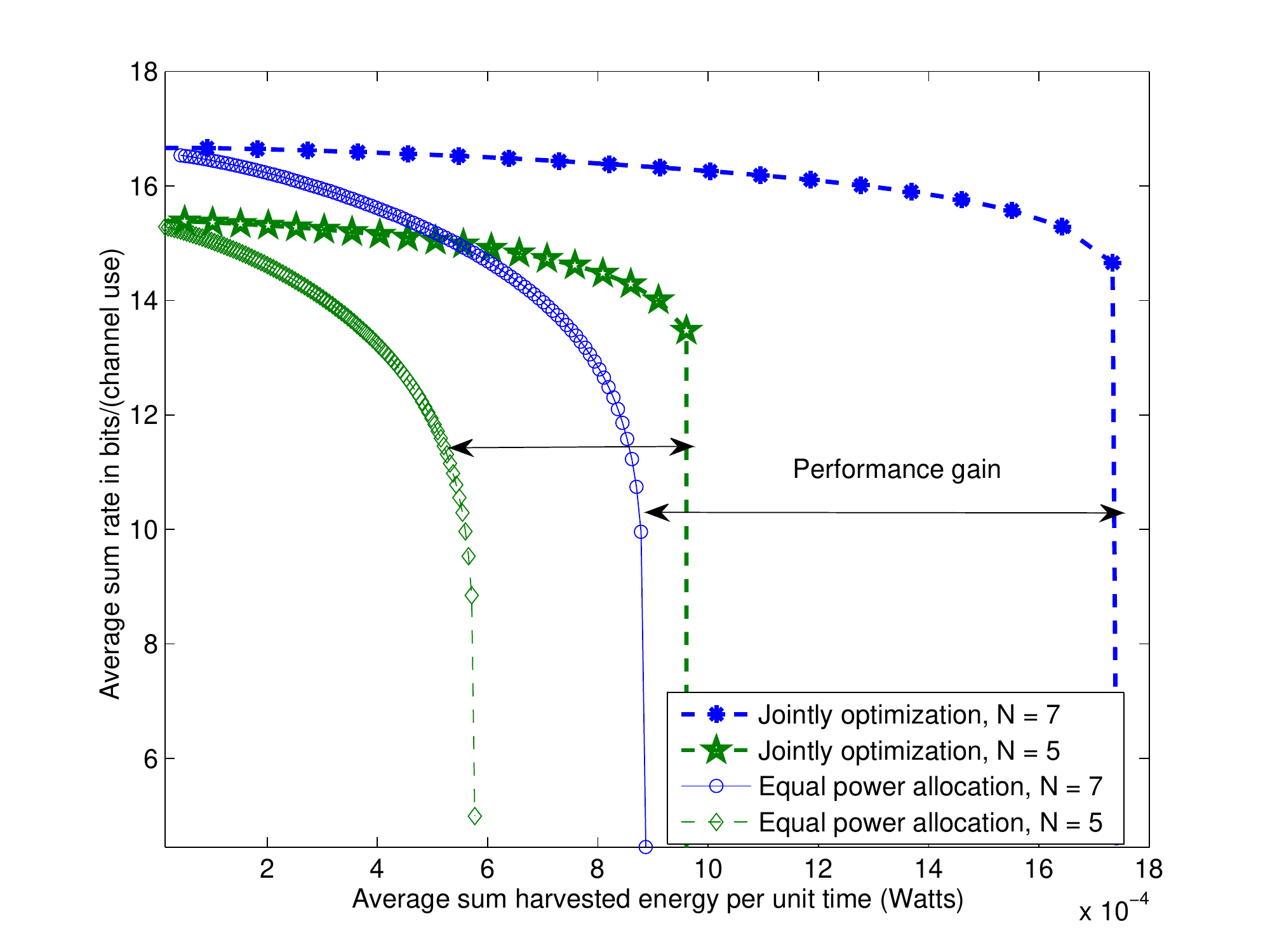}
\caption{Average sum rate versus average sum harvested energy of the MT scheme with joint power allocation and user selection for different numbers of UTs.}
\label{fig:UBvsFixedRice}
\end{figure}
\subsection*{Simulation}
Next, we perform simulations to verify the performance of the proposed jointly optimal power allocation and user scheduling algorithm. We adopt the same setup as in Chapter 2. The maximum average transmit power $P_{\mathrm{ave}}$ is $10$ Watt and the maximum instantaneous transmit power  $P_{\max}$ is set to $46$ dBm. Figure \ref{fig:UBvsFixedRice} shows the  R-E regions for the proposed optimal resource allocation scheme with joint user selection and power allocation for different power allocation schemes and different users. For comparison, we also show the performance of the fixed power resource allocation scheme proposed in Chapter 2. It can be observed that  the average achievable rate and the average harvested energy increases with the number of users. In fact, the joint power allocation and user selection can effectively exploit the channel fluctuations to improve the system performance.  Besides, a large harvested power gain can be achieved
 by the proposed joint optimization scheme over the fixed power allocation scheme.  The power allocation variables provide more degrees of freedom across different time slots in resource allocation. Specifically, compared to fixed power transmission, a larger power is transmitted to improve the system performance when the channel condition is good. Also, a smaller transmit power is allocated when the channel is in deep fading which preserves the energy of transmitter  to exploit the good channel conditions in the future.

\chapter{Conclusion}
In this thesis, we have proposed optimal multiuser scheduling schemes for SWIPT systems considering different notions of fairness in resource allocation. The designed schemes enable the control of the tradeoff between the average sum rate and the average amount of sum harvested energy. Our results reveal that for the maximization of the system sum rate with or without fairness constraints, the optimal scheduling algorithm  requires only causal  instantaneous and statistical channel knowledge. Simulation results revealed that
substantial performance gains can be achieved by the proposed optimization framework compared to existing suboptimal scheduling schemes. Besides, joint user selection and power allocation was also investigated for the considered SWIPT system. The obtained solution revealed that the proposed optimal resource allocation scheme improves the performance and enlarges the feasible R-E region compared to the scheme with fixed AP transmit power. Further investigation on the impact of imperfect CSI for resource allocation   is left for future work.


\bibliographystyle{IEEEtran}
\bibliography{literature}

\appendix
\chapter{Proof of Optimality of Binary Relaxation}
\section{Optimality of Binary Relaxation for Optimum Multiuser Scheduling Schemes}
\label{AppendixA1.1}

We prove that the optimal solution of the problem in \eqref{eqn:selection_metric} with the relaxed constraint, $0 \leq q_n (i) \leq 1$, selects the boundary values of $q_n (i)$, i.e., $0$ or $1$. Therefore, the binary relaxation does not change the solution of the problem.

If one of the $q_n (i)$ adopts a non-binary value in the optimal solution, there has to be at least one other non-binary selection variable in the same time slot $i$. We assume that the indices of the non-binary selection variables are $n^{'}$ and $n^{''}$ in the $i$-th time slot. Then, for the optimization problem corresponding to the MT scheduling scheme in \eqref{eqn:selection_metric} we obtain $\alpha_n (i) = 0 \hspace{5pt} \forall n$ from \eqref{eqn:slackness_conditions_MT1} and $\beta_{n^{'}} (i) = 0$ and $\beta_{n^{''}} (i) = 0$ from \eqref{eqn:slackness_conditions_MT2}. By substituting these values into \eqref{eqn:partial_d_MT}, we obtain
\begin{subequations}
\begin{alignat}{3}
& T \left( \lambda (i) \right) = \Lambda_{n^{'}}(i), \\
& T \left( \lambda (i) \right) = \Lambda_{n^{''}}(i).
\end{alignat}
\end{subequations}

From (A.1a) and (A.1b), it follows that $\Lambda_{n^{'}}(i) = \Lambda_{n^{''}}(i)$. However, due to the randomness of the time-continuous channel gains, $\text{Pr} \{ \Lambda_{n^{'}}(i) = \Lambda_{n^{''}}(i) \} = 0$, where $\text{Pr} \{ \cdot \} $ denotes probability. Therefore, the optimal $q_n (i) \in \{ 0, 1\}$ $\forall n, i$ and \eqref{eqn:selection_metric} is the optimal selection policy.

The tightness of binary constraint relaxation for the PF scheduling and the ET scheduling can be proved by following a similar approach as for the MT scheduling.

%
%
\end{document}

%% file: Macros.tex
\newcommand*{\argmin}{\ensuremath{\mathop{\mathrm{arg\,min}}}}
\newcommand*{\argmax}{\ensuremath{\mathop{\mathrm{arg\,max}}}}
\DeclareMathOperator{\mino}{minimize}

%% file: 0_Title.tex
\newlength{\logoheight}\setlength{\logoheight}{20mm}
\newlength{\logomargin}\setlength{\logomargin}{35mm}

\addcontentsline{toc}{chapter}{\titlename}
\begin{titlepage}
\begin{tikzpicture}[remember picture, overlay]
  \begin{scope}[every node/.style={text badly centered,text width=1.1\textwidth}]
    \path (current page.north)
      +(0mm,-40mm)  node[font=\Large] {\Thesis}
      +(0mm,-60mm)  node[font=\huge\bfseries,minimum height=5cm] {\Title}
      +(0mm,-95mm)  node[font=\large] {\Author}
      +(0mm,-135mm) node[font=\large] {
        \textbf{Lehrstuhl f\"{u}r Digitale \"{U}bertragung}\\
        Prof. Dr.-Ing. Robert Schober\\
        Universit\"{a}t Erlangen-N\"{u}rnberg\\
      }
      +(0mm,-175mm) node {
        \begin{tabular}{ll}
          Supervisors: & Rania Morsi, M.Sc.\\
                    & Derrick Wing Kwan Ng, Ph.D.\\
        \end{tabular}
      }
      +(0mm,-225mm) node {\Date}
    ; 
  \end{scope}
  \node[shift={(+\logomargin,1.5\logoheight)},anchor=west] at (current page.south west){%
    \includegraphics[height=\logoheight]{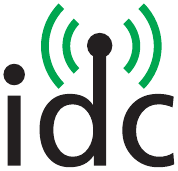}%
  };
  \node[shift={(-\logomargin,1.5\logoheight)},anchor=east] at (current page.south east){%
    \includegraphics[height=\logoheight]{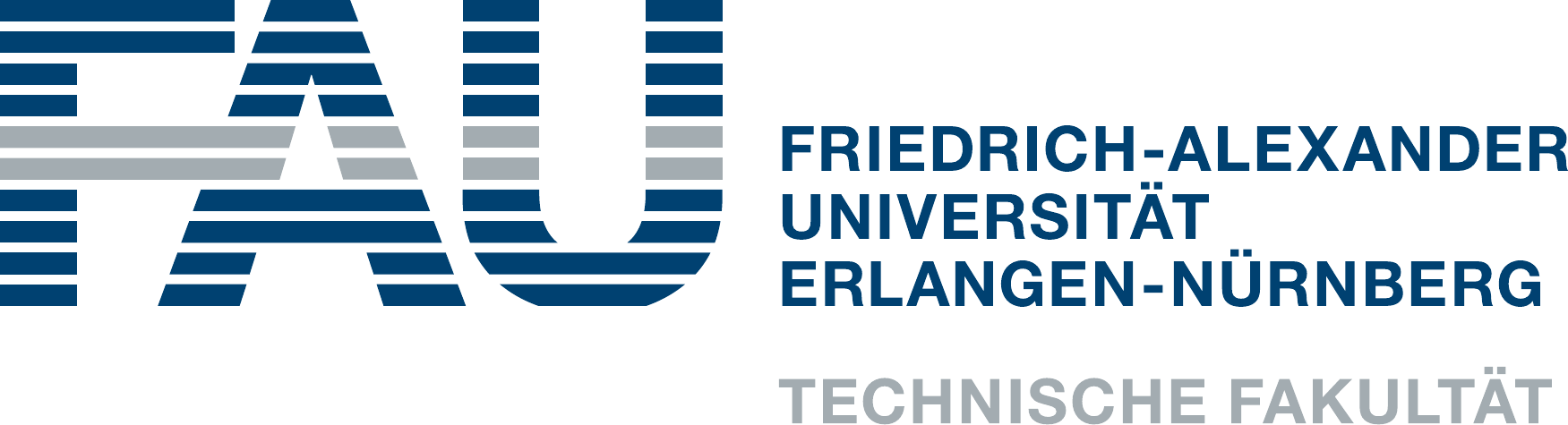}%
  };
\end{tikzpicture}
\end{titlepage}

%% file: 0_Abstract.tex

In this thesis, we study the downlink multiuser scheduling and power allocation problem for systems with simultaneous wireless information and power transfer (SWIPT).  In the first part of the thesis, we focus on multiuser scheduling.   We design optimal scheduling algorithms that maximize the long-term average  system throughput under different fairness requirements, such as  proportional fairness and equal throughput fairness. In particular, the algorithm designs are formulated as  non-convex optimization problems which take into account the minimum required  average sum harvested energy in the system.   The problems are solved by using convex optimization techniques and the proposed optimization framework reveals the tradeoff between the long-term average system throughput and the sum harvested energy in multiuser systems with fairness constraints.  Simulation results demonstrate that
substantial performance gains can be achieved by the proposed optimization framework compared to existing suboptimal scheduling algorithms from the literature. In the second part of the thesis, we investigate the joint user scheduling and power allocation algorithm design for SWIPT systems. The algorithm design is formulated as a non-convex optimization problem which  maximizes the achievable rate subject to a minimum required average power transfer. Subsequently, the non-convex optimization problem is reformulated by big-M method  which can be solved optimally. Furthermore, we show that joint power allocation and user scheduling is an efficient
way to enlarge the feasible trade-off region for improving the system performance in
terms of achievable data rate and harvested energy. 

%% file: 0_Glossary.tex
\newcommand{\glossaryfirstcolumnlength}{\hspace{6.1em}}
\addsec{\operatorsname}
\begin{symbollist}{\glossaryfirstcolumnlength}
  \sym{$\argmax$}{Argument of the maximum value}
  \sym{$\argmin$}{Argument of the minimum value}
  \sym{$\text{arg order}$}{Argument of the $j^{th}$ ascending order value}
  \sym{$\text{Pr} \{ \cdot \} $}{Probability of an event}
  \sym{$\nabla_x$}{Gradient}
\end{symbollist}
\addsec{\symbolsname}
\begin{symbollist}{\glossaryfirstcolumnlength}
   \sym{$C_n (i)$}{Achievable rate of user $n$ in time slot $i$}
   \sym{$\bar{C}_n$}{Average rate of user $n$}
   \sym{$h_n (i)$}{Channel power gain from the access point to user $n$ in time slot $i$}
   \sym{$i$}{Time slot index}
   \sym{$j$}{Selection order}
   \sym{$L$}{Lagrangian function}
  \sym{$n$}{User index}
  \sym{$N$}{Number of users in the system}
  \sym{$P$}{Constant transmit power of the access point}
  \sym{$P_n (i)$}{Power allocated to user $n$ from the access point in the downlink in time slot $i$}
   \sym{$q_n (i)$}{User selection variable for user $n$ in time slot $i$}
   \sym{$Q_n (i)$}{Energy harvested by user $n$ in time slot $i$}
   \sym{$Q_{\mathrm{req}}$}{Required amount of average sum harvested energy}
   \sym{$\bar{Q}_{\mathrm{sum}}$}{Actual average sum harvested energy}
  \sym{$r$}{Minimum rate per user}
  \sym{$\bar{R}_{\mathrm{sum}}$}{Average sum rate}
  \sym{$T$}{Total number of time slots}
\sym{$\gamma_n$}{Lagrange multiplier which ensures proportional fairness}
\sym{$\Gamma$}{Step size for Lagrange multiplier $\gamma_n$ in the gradient algorithm}
\sym{$\zeta$}{Step size for Lagrange multiplier $\theta_n$ in the gradient algorithm}
\sym{$\eta$}{Proportion of time when only energy harvesting is performed}
\sym{$\theta_n$}{Lagrange multiplier which ensures equal throughput fairness}
\sym{$\Theta$}{Step size for Lagrange multiplier $\nu$ in the gradient algorithm}
\sym{$\lambda (i)$}{Lagrange multiplier which corresponds to the constraint that only one user is chosen}
\sym{$\Lambda_n (i)$}{Selection metric of user $n$ in time slot $i$}
\sym{$\nu$}{Lagrange multiplier which corresponds to the constraint on the sum harvested energy}
\sym{$\xi_n$}{RF-to-DC conversion efficiency of receiver $n$}
\sym{$\sigma^2$}{Noise power}
\sym{$\Omega_n$}{Mean channel power gain of user $n$ in time slot $i$}
\end{symbollist}
\printglossaries
\clearpage

%% file: Main_Document_updated.bbl
\begin{thebibliography}{10}
\providecommand{\url}[1]{#1}
\csname url@samestyle\endcsname
\providecommand{\newblock}{\relax}
\providecommand{\bibinfo}[2]{#2}
\providecommand{\BIBentrySTDinterwordspacing}{\spaceskip=0pt\relax}
\providecommand{\BIBentryALTinterwordstretchfactor}{4}
\providecommand{\BIBentryALTinterwordspacing}{\spaceskip=\fontdimen2\font plus
\BIBentryALTinterwordstretchfactor\fontdimen3\font minus
  \fontdimen4\font\relax}
\providecommand{\BIBforeignlanguage}[2]{{%
\expandafter\ifx\csname l@#1\endcsname\relax
\typeout{** WARNING: IEEEtran.bst: No hyphenation pattern has been}%
\typeout{** loaded for the language `#1'. Using the pattern for}%
\typeout{** the default language instead.}%
\else
\language=\csname l@#1\endcsname
\fi
#2}}
\providecommand{\BIBdecl}{\relax}
\BIBdecl

\bibitem{Powercast}
\BIBentryALTinterwordspacing
{Powercast Coporation}, ``{RF} {E}nergy {H}arvesting and {W}ireless {P}ower for
  {L}ow-{P}ower {A}pplications,'' 2011. [Online]. Available:
  \url{http://www.mouser.com/pdfdocs/Powercast-Overview-2011-01-25.pdf}
\BIBentrySTDinterwordspacing

\bibitem{Krikidis2014}
I.~Krikidis, S.~Timotheou, S.~Nikolaou, G.~Zheng, D.~W.~K. Ng, and R.~Schober,
  ``Simultaneous {W}ireless {I}nformation and {P}ower {T}ransfer in {M}odern
  {C}ommunication {S}ystems,'' \emph{IEEE Commun. Mag.}, vol.~52, no.~11, pp.
  104--110, Nov. 2014.

\bibitem{Ding2014}
Z.~Ding, C.~Zhong, D.~W.~K. Ng, M.~Peng, H.~A. Suraweera, R.~Schober, and H.~V.
  Poor, ``Application of smart {A}ntenna {T}echnologies in {S}imultaneous
  {W}ireless {I}nformation and {P}ower {T}ransfer,'' 2015, to appear in the
  \emph{ IEEE Commun. Mag.}

\bibitem{Varshney2010}
L.~Varshney, ``Transporting {I}nformation and {E}nergy {S}imultaneously,''
  \emph{IEEE Intern. Symp. Inform. Theory (ISIT)}, pp. 1612--1616, Jul. 2008.

\bibitem{Grover2008}
P.~Grover and A.~Sahai, ``Shannon {M}eets {T}esla: Wireless {I}nformation and
  {P}ower {T}ransfer,'' \emph{IEEE Intern. Symp. Inform. Theory (ISIT)}, pp.
  2363--2367, June 2010.

\bibitem{Zhang2013}
R.~Zhang and C.~K. Ho, ``{MIMO} {B}roadcasting for {S}imultaneous {W}ireless
  {I}nformation and {P}ower {T}ransfer,'' \emph{IEEE Trans. Wireless Commun.},
  vol.~12, no.~5, pp. 1989--2001, May 2013.

\bibitem{Zhou2013}
X.~Zhou, R.~Zhang, and C.~K. Ho, ``Wireless {I}nformation and {P}ower
  {T}ransfer: {D}esign and {R}ate-{E}nergy {T}radeoff,'' \emph{IEEE Trans. on
  Commun.}, vol.~61, no.~11, pp. 4754--4767, November 2013.

\bibitem{CN:Kwan_globecom2013}
D.~W.~K. Ng and R.~Schober, ``{Resource Allocation for Secure Communication in
  Systems with Wireless Information and Power Transfer},'' in \emph{Proc. IEEE
  Global Telecommun. Conf.}, Dec. 2013.

\bibitem{CN:Eurosip_SWIPT}
------, ``{Spectral Efficient Optimization in OFDM Systems With Wireless
  Information and Power Transfer},'' in \emph{21st European Signal Process.
  Conf. (EUSIPCO)}, Sep. 2013, pp. 1--5.

\bibitem{CN:ICC_WIPT_Kwan}
D.~W.~K. Ng, E.~S. Lo, and R.~Schober, ``{Energy-Efficient Power Allocation in
  OFDM Systems with Wireless Information and Power Transfer},'' in \emph{Proc.
  IEEE Intern. Commun. Conf.}, Jun. 2013, pp. 4125--4130.

\bibitem{JR:Kwan_secure_imperfect}
------, ``{Robust Beamforming for Secure Communication in Systems with Wireless
  Information and Power Transfer},'' \emph{IEEE Trans. Wireless Commun.},
  vol.~13, pp. 4599--4615, Aug. 2014.

\bibitem{JR:WIPT_fullpaper}
------, ``{Wireless Information and Power Transfer: Energy Efficiency
  Optimization in OFDMA Systems},'' \emph{IEEE Trans. Wireless Commun.},
  vol.~12, pp. 6352 -- 6370, Dec. 2013.

\bibitem{CN:WCNC_WIPT}
------, ``{Energy-Efficient Resource Allocation in Multiuser OFDM Systems with
  Wireless Information and Power Transfer},'' in \emph{Proc. IEEE Wireless
  Commun. and Netw. Conf.}, 2013.

\bibitem{CN:kwan_vicky}
S.~Leng, D.~W.~K. Ng, and R.~Schober, ``{Power Efficient and Secure Multiuser
  Communication Systems with Wireless Information and Power Transfer},'' in
  \emph{Proc. IEEE Intern. Commun. Conf.}, Jun. 2014.

\bibitem{CN:Kwan_PIMRC2013}
D.~W.~K. Ng, L.~Xiang, and R.~Schober, ``{Multi-Objective Beamforming for
  Secure Communication in Systems with Wireless Information and Power
  Transfer},'' in \emph{Proc. IEEE Sympos. on Personal, Indoor and Mobile Radio
  Commun.}, Sep. 2013.

\bibitem{CN:Multicast_SWIPT}
D.~W.~K. Ng, R.~Schober, and H.~Alnuweiri, ``{Secure Layered Transmission in
  Multicast Systems With Wireless Information and Power Transfer},'' in
  \emph{Proc. IEEE Intern. Commun. Conf.}, Jun. 2014, pp. 5389--5395.

\bibitem{CN:Kwan_globecom2014}
D.~W.~K. Ng and R.~Schober, ``{Resource Allocation for Coordinated Multipoint
  Networks With Wireless Information and Power Transfer},'' in \emph{Proc. IEEE
  Global Telecommun. Conf.}, Dec. 2014, pp. 4281--4287.

\bibitem{CN:Maryna_2015}
M.~Chynonova, R.~Morsi, D.~W.~K. Ng, and R.~Schober, ``{Optimal Multiuser
  Scheduling Schemes for Simultaneous Wireless Information and Power
  Transfer},'' in \emph{Proc. 23rd European Signal Process. Conf. (EUSIPCO)},
  2015.

\bibitem{CN:tao_2015}
Q.~Wu, M.~Tao, D.~W.~K. Ng, W.~Chen, and R.~Schober, ``{Energy-Efficient
  Transmission for Wireless Powered Multiuser Communication Networks},'' in
  \emph{Proc. IEEE Intern. Commun. Conf.}, Jun. 2015.

\bibitem{JR:MOOP_SWIPT}
\BIBentryALTinterwordspacing
D.~W.~K. Ng, E.~S. Lo, and R.~Schober, ``{Multi-Objective Resource Allocation
  for Secure Communication in Cognitive Radio Networks with Wireless
  Information and Power Transfer},'' \emph{CoRR}, 2014. [Online]. Available:
  \url{http://arxiv.org/abs/1403.0054}
\BIBentrySTDinterwordspacing

\bibitem{JR:rui_zhang}
L.~Liu, R.~Zhang, and K.-C. Chua, ``{Secrecy Wireless Information and Power
  Transfer with MISO Beamforming},'' \emph{IEEE Trans. Signal Process.},
  vol.~62, pp. 1850--1863, Apr. 2014.

\bibitem{JR:Kwan_SEC_DAS}
\BIBentryALTinterwordspacing
D.~W.~K. Ng and R.~Schober, ``{Secure and Green {SWIPT} in Distributed Antenna
  Networks with Limited Backhaul Capacity},'' \emph{CoRR}, 2014. [Online].
  Available: \url{http://arxiv.org/abs/1410.3065}
\BIBentrySTDinterwordspacing

\bibitem{CN:PHY_SEC_max_min}
D.~Ng and R.~Schober, ``{Max-Min Fair Wireless Energy Transfer for Secure
  Multiuser Communication Systems},'' in \emph{IEEE Inf. Theory Workshop
  (ITW)}, Nov. 2014, pp. 326--330.

\bibitem{Xu2013}
J.~Xu, L.~Liu, and R.~Zhang, ``Multiuser {B}eamforming for {S}imultaneous
  {W}ireless {I}nformation and {P}ower {T}ransfer,'' \emph{IEEE Intern. Conf.
  on Acoustics, Speech and Signal Processing (ICASSP)}, pp. 4754--4758, May
  2013.

\bibitem{Park2013}
J.~Park and B.~Clerckx, ``Joint {W}ireless {I}nformation and {E}nergy
  {T}ransfer in a {T}wo-{U}ser {MIMO} {I}nterference {C}hannel,'' \emph{IEEE
  Trans. Wireless Commun.}, vol.~12, no.~8, pp. 4210--4221, Aug. 2013.

\bibitem{Morsi2014}
R.~Morsi, D.~Michalopoulos, and R.~Schober, ``Multi-{U}ser {S}cheduling
  {S}chemes for {S}imultaneous {W}ireless {I}nformation and {P}ower
  {T}ransfer,'' \emph{Proc. IEEE Intern. Commun. Conf.}, pp. 4994--4999, Jun.

\bibitem{LiuZahng2014}
L.~Liu, , R.~Zhang, and K.~C. Chua, ``Multi-{A}ntenna {W}ireless {P}owered
  {C}ommunication with {E}nergy {B}eamforming,'' \emph{ArXiv e-prints}, 2014,
  arxiv:1312.1450.

\bibitem{Huang2014}
K.~Huang and V.~K.~N. Lau, ``Enabling {W}ireless {P}ower {T}ransfer in
  {C}ellular {N}etworks: Architecture, {M}odeling and {D}eployment,''
  \emph{IEEE Trans. Wireless Commun.}, vol.~13, no.~2, pp. 902--912, Feb. 2014.

\bibitem{Shi2011}
Y.~Shi, L.~Xie, Y.~Hou, and H.~Sherali, ``On {R}enewable {S}ensor {N}etworks
  with {W}ireless {E}nergy {T}ransfer,'' \emph{Proc. IEEE INFOCOM}, pp.
  1350--1358, Apr. 2011.

\bibitem{Lee2013}
S.~H. Lee, R.~Zhang, and K.~Huang, ``Opportunistic {W}ireless {E}nergy
  {H}arvesting in {C}ognitive {R}adio {N}etworks,'' \emph{IEEE Trans. Wireless
  Commun.}, vol.~12, no.~9, pp. 4788--4799, Sep. 2013.

\bibitem{Ju2014}
H.~Ju and R.~Zhang, ``Throughput {M}aximization in {W}ireless {P}owered
  {C}ommunication {N}etworks,'' \emph{IEEE Trans. Wireless Commun.}, vol.~13,
  no.~1, pp. 418--428, January 2014.

\bibitem{Huang2013}
C.~Huang, R.~Zhang, and S.~Cui, ``Throughput {M}aximization for the {G}aussian
  {R}elay {C}hannel with {E}nergy {H}arvesting {C}onstraints,'' \emph{IEEE J.
  Select. Areas Commun.}, vol.~31, no.~8, pp. 1469--1479, Aug. 2013.

\bibitem{Gurakan2012}
B.~Gurakan, O.~Ozel, J.~Yang, and S.~Ulukus, ``Energy {C}ooperation in {E}nergy
  {H}arvesting {W}ireless {C}ommunications,'' \emph{Proc. IEEE Intern. Sympos.
  on Inf. Theory}, Jul. 2012.

\bibitem{JuZahng2014}
H.~Ju and R.~Zhang, ``User {C}ooperation in {W}ireless {P}owered
  {C}ommunication {N}etworks,'' \emph{ArXiv e-prints}, 2014, arxiv:1403.7123.

\bibitem{ZahngJu2014}
{H. Ju} and R.~Zhang, ``Optimal {R}esource {A}llocation in {F}ull-{D}uplex
  {W}oreless-{P}owered {C}ommunication {N}etwork,'' \emph{ArXiv e-prints},
  2014, arxiv:1403.2580.

\bibitem{Rappaport}
T.~S. Rappaport, \emph{{W}ireless {C}ommunications: {P}rinciples and
  {P}ractice}, 2nd~ed.\hskip 1em plus 0.5em minus 0.4em\relax Prentice {H}all,
  2002.

\bibitem{Boyd2004}
S.~Boyd and L.~Vandenberghe, \emph{Convex {O}ptimization}.\hskip 1em plus 0.5em
  minus 0.4em\relax New York, NY, USA: Cambridge University Press, 2004.

\end{thebibliography}
